\documentclass[11pt]{JHEP3}

\usepackage{amsmath}
\usepackage{amssymb}
\usepackage{graphicx}
\usepackage{dsfont}
\usepackage{overpic}
\usepackage{cite}
\usepackage{slashed}

\newcommand{\beq}{\begin{equation}}
\newcommand{\eeq}{\end{equation}}
\newcommand{\beqa}{\begin{eqnarray}}
\newcommand{\eeqa}{\end{eqnarray}}

\title{Infrared regularization with vector mesons and baryons}

\author{Peter Christian Bruns \\ Universit\"at Bonn,
Helmholtz--Institut f\"ur Strahlen-- und Kernphysik (Theorie),\\
D-53115 Bonn, Germany \\ E-mail: \email{bruns@itkp.uni-bonn.de}}
\author{Ulf-G. Mei\ss ner \\ Universit\"at Bonn,
Helmholtz--Institut f\"ur Strahlen-- und Kernphysik (Theorie)\\
and Bethe Center for Theoretical Physics,
D-53115 Bonn, Germany \\ and \\ Forschungszentrum J\"ulich, Institut f\"ur Kernphysik 
(Theorie) \\  and J\"ulich Center for Hadron Physics,
D-52425 J\"ulich, Germany \\ E-mail: \email{meissner@itkp.uni-bonn.de}}

\abstract{We extend the method of infrared regularization to spin-1 fields coupled to
baryons. As an application, we discuss the axial form factor of the nucleon.}


\preprint{FZJ-IKP-TH-2008-13 \quad HISKP-TH-08/14}

\begin{document}

\section{Introduction}

In a recent work \cite{BM04}, the scheme of infrared regularization (developed
in its original form by Becher and Leutwyler \cite{BL}) was extended to the
case of explicit meson resonances interacting with soft pions, as e.g. the
first step towards a systematic inclusion of vector mesons in the meson sector
of Chiral Perturbation Theory (ChPT). 
It is the aim of this article to provide a further extension of infrared
regularization to 
the situation where baryons as well as vector mesons are present. This can 
be considered as a synthesis of the results in \cite{BM04} and \cite{BL}.
Such an extension is not only of interest in itself, but can be applied
to a plethora of observables, where vector and axial-vector mesons are
known to play an important role, such as the electroweak form factors
of the nucleon. As one example we mention the contribution of the $\pi\rho$
loop to the strangeness form factors of the nucleon~\cite{Meissner:1997qt,Hammer:1997vt}.

Infrared regularization (IR) is a solution to the following problem. The presence
of mass scales which must be considered as 'heavy' compared to the masses and
momenta of the soft pions will in general mess up the usual power counting
rules of ChPT by which the perturbation series of the effective theory is 
ordered \cite{Wein,GL84,GL85}. This observation was first made when baryons
were incorporated in the framework of ChPT \cite{GSS} (for a recent 
review on baryon ChPT we refer to \cite{Bernard:2007zu}). The procedure of
infrared regularization separates the (dimensionally regularized) one-loop
graphs of baryon ChPT into a part which stems from the soft pion contribution
and a part generated from loop momenta close to the 'heavy' scale. The latter 
portion of the loop graph, called the 'regular' part, will usually not be in 
accord with the low-energy power counting, but can always be absorbed in local
terms derived from the effective Lagrangian. It is therefore dropped from the 
loop graph, and only the first part, called the infrared singular part of the 
loop integral, is kept. Though both the vector mesons and the baryons interact
as 'heavy' particles with the pions, the power counting is different for the
two species, at least for the kinds of Feynman graphs we consider in this
work. There, the vector mesons appear only as internal lines with small
momenta far from their mass shell, so that the resonance propagator is counted 
as $O(q^{0})$ (where $q$ indicates some small momentum scale or Goldstone
boson mass), while the baryon propagator is counted as $O(q^{-1})$, since the
baryon is pushed from its mass shell only by a small amount due to its
interaction with the soft pions and vector mesons. In the graphs we treat
here, only one single baryon is present, with the baryon line running through
the diagram undergoing only soft interactions. The number of the virtual
vector mesons, however, is not fixed. The pion propagator is counted as
$O(q^{-2})$, as usual, both the pion momentum and the pion mass being of
$O(q)$. Appropriate powers of $q$ are also assigned to vertices from the 
effective Lagrangian. Finally, the measure of every $d$-dimensional loop
integration is booked as $O(q^{d})$. This counting scheme applies, of course,
to tree graphs, but also to the infrared singular, or soft, parts of the loop
graphs. The regular parts of the loop graphs are not guaranteed to obey the
power counting rules. These general remarks will be exemplified in the
following sections. 

Before working out the case where both baryons and vector
mesons appear in a Feynman diagram, we will briefly  review the scheme of
infrared regularization for loop integrals where only one heavy scale shows
up. This will not only serve to give a unified presentation of the method, but
also provide some results needed for an application of the general scheme to 
the axial form factor of the nucleon. Before starting with the presentation of 
the formalism, let us mention that the loop integrals studied in this work
have also been treated, using a different regularization scheme which is in 
some respect complementary to the one used here, in Ref.~\cite{Mainz03}.

\section{IR regularization in the pion-nucleon system} \label{sec:PiN}

When only pions and nucleons are treated as explicit fields of the effective
theory, the fundamental loop integral one has to consider is
\begin{equation} \label{eq:IMB}
I_{MB}(p^2) = \int\frac{d^{d}l}{(2\pi)^{d}}\frac{i}{((p-l)^2-m^2)(l^2-M^2)} .
\end{equation}
Here, $M$ is the pion mass (being of chiral order $O(q)$) and $m$ is 
the nucleon mass. Applying the low-energy power counting scheme outlined 
in the introduction, one would assign a chiral order of $q^{d-3}$ to this 
integral. We will see in a moment that only an appropriately extracted 
low-energy part of $I_{MB}$ will obey this power counting requirement. 

All the other pion-nucleon loop integrals are either only trivially 
modified by the infrared regularization scheme, or they can be 
derived from eq.~(\ref{eq:IMB}) (see sec.~6 of \cite{BL}). For example, 
the scalar tadpole integral containing only the pion propagator is not 
modified at all, as there is no 'hard momentum' structure present that could 
lead to a nonvanishing regular part of this integral. Thus we have
\begin{displaymath}
I_{M}^{IR} = I_{M} = \int\frac{d^{d}l}{(2\pi)^{d}}\frac{i}{l^{2}-M^{2}} ,
\end{displaymath}
and a direct calculation gives
\begin{equation} \label{eq:IM}
I_{M}^{IR} = \frac{\Gamma(1-\frac{d}{2})}{(4\pi)^{\frac{d}{2}}}M^{d-2} .
\end{equation}
Note the typical structure of the $d$-dependent power of the pion mass. 
For arbitrary values of the dimension parameter $d$, $I_{M}^{IR}$ is in 
general proportional to fractional powers of $M^{2}$, and will even diverge 
in the so-called chiral limit where the quark masses $m_{u},m_{d}$ go to zero 
(so that also $M\rightarrow 0$) for small enough $d$ (we mostly consider the
two--flavor case here but the method can be trivially extended to include also
strange quarks). Such terms will never occur in the regular parts of the 
loop integrals which stem from the high-momentum region of the integration: 
those parts are always expandable in the pion mass. The baryon tadpole
integral, e.g., is
\begin{displaymath}
I_{B} = \int\frac{d^{d}l}{(2\pi)^{d}}\frac{i}{l^{2}-m^{2}},
\end{displaymath}
and is trivially expandable in $M^{2}$, since it has no pion mass 
dependence at all. Therefore one has $I_{B}^{IR}=0$. These remarks may 
serve to explain the terminology 'infrared singular' vs. 'regular'. 

Returning to eq.~(\ref{eq:IMB}), we must extract the part of this 
integral that is proportional to some $d$-dependent power of $M$, 
like in eq.~(\ref{eq:IM}). One can think of this extraction prescription 
as an operational definition of infrared regularization.
The method is explained in full generality in \cite{BL}. Here, 
we concentrate on the case with on-shell momentum $p$, i.e. $p^{2}=m^{2}$. 
This shows all the features we need for the demonstration, and also yields 
the result we will use in our application of the scheme in sec.~\ref{sec:PiNV}.
We introduce a Feynman parameter integration in the usual way:
\begin{equation} \label{eq:intlz}
I_{MB} =
\int\frac{d^{d}l}{(2\pi)^{d}}\int_{0}^{1}\frac{dz}{[((p-l)^{2}-m^{2})z 
+ (l^{2}-M^{2})(1-z)]^{2}}.
\end{equation} 
Performing the standard steps, we find for $p^{2}=m^{2}$:
\begin{equation}
I_{MB} = -m^{d-4}\frac{\Gamma(2-\frac{d}{2})}{(4\pi)^{\frac{d}{2}}}
\int_{0}^{1}\frac{dz}{(z^{2}-\alpha z + \alpha)^{2-\frac{d}{2}}} , 
\end{equation}
where we have defined $\alpha = M^{2}/m^{2}$ (note the difference to 
ref.\cite{BL}, where this letter is reserved for $\alpha_{\mathrm{BL}}= M/m$).
Fractional powers of the small variable $\alpha$ will be produced near $z=0$: 
there, the integrand is approximately $(\alpha)^{\frac{d}{2}-2}$. For small 
enough $d$, there would be an infrared singularity for $M\rightarrow 0$ 
located in parameter space at $z=0$. It can already be seen from
eq.~(\ref{eq:intlz}) 
that the parameter region near $z=0$ is associated with the low-energy portion 
of the integral: In this region, only the 'soft' pion propagator is weighted 
in the loop integration, while the hard momentum structure of the nucleon 
propagator dominates near $z=1$.
The extraction of the part of the integral proportional to $d$-dependent
powers of $\alpha$ now proceeds as follows: the parameter integration is 
split into two parts like
\begin{equation} \label{eq:split}
\int_{0}^{1} = \int_{0}^{\infty} - \int_{1}^{\infty}.
\end{equation}
We will first show that the second integral on the r.h.s. is 
regular, i.e. expandable, in the variable $\alpha$. This is easy to see, 
because for $z\geq 1$, the integrand can be expanded like
\begin{equation} \label{eq:expintR}
(z^{2}-\alpha z + \alpha)^{\frac{d}{2}-2} 
=
z^{d-4}\sum_{k=0}^{\infty}\frac{\Gamma(2-\frac{d}{2}+k)}{\Gamma(2-\frac{d}{2})}
\frac{\alpha^{k}}{k!}\biggl(\frac{z-1}{z^{2}}\biggr)^{k}.
\end{equation}
Interchanging integration and summation (which is a valid operation at least 
for some range of $d$), one gets for the regular part
\begin{equation} \label{eq:R}
R \equiv m^{d-4}\frac{\Gamma(2-\frac{d}{2})}{(4\pi)^{\frac{d}{2}}}
\int_{1}^{\infty}\frac{dz}{(z^{2}-\alpha z + \alpha)^{2-\frac{d}{2}}} 
= m^{d-4}\frac{\Gamma(2-\frac{d}{2})}{(3-d)(4\pi)^{\frac{d}{2}}} + O(\alpha).
\end{equation}
At this point we should make the remark that the extension of the parameter 
integration to infinity will lead to divergences as $d$ increases. The
infrared singular or regular parts are then defined as follows: The parameter 
integrals are computed for the range of $d$ where they are well-defined, and 
the result will be continued analytically to arbitrary values of $d$. 
This amounts to the suppression of power divergences of the parameter
integrals, which have nothing to do with the infrared singularity at $z=0$, 
and will be cancelled anyway on the r.h.s. of eq.~(\ref{eq:split}).

Next we must show that the first term on the r.h.s. of eq.~(\ref{eq:split}) 
is proportional to a $d$-dependent power of $\alpha$. To see this, 
we substitute $z=\sqrt{\alpha}y$ to get
\begin{displaymath} 
\int_{0}^{\infty} \frac{dz}{(z^{2}-\alpha z + \alpha)^{2-\frac{d}{2}}} 
= \sqrt{\alpha}^{d-3}\int_{0}^{\infty}\frac{dy}{(y^{2}-\sqrt{\alpha} y + 1)^{2-\frac{d}{2}}}. 
\end{displaymath}   
The remaining integral on the r.h.s. will not produce $d$-dependent powers 
of $\alpha$, since the integrand can be expanded in $\sqrt{\alpha}$ similar 
to eq.~(\ref{eq:expintR}). Thus we have found that the parameter integral from 
zero to infinity is proportional to a $d$-dependent power of $\alpha$. 
The infrared singular part $I_{MB}^{IR}$ of the loop integral therefore equals
\begin{eqnarray} \label{eq:IMBIR}
I_{MB}^{IR} &=& -m^{d-4}\frac{\Gamma(2-\frac{d}{2})}{(4\pi)^{\frac{d}{2}}}
\int_{0}^{\infty}\frac{dz}{(z^{2}-\alpha z + \alpha)^{2-\frac{d}{2}}} \nonumber \\  
&=& -\frac{m^{d-4}\sqrt{\alpha}^{d-3}}{2(4\pi)^{\frac{d}{2}}}
\sum_{k=0}^{\infty}\frac{\sqrt{\alpha}^{k}}{k!}\Gamma
\biggl(\frac{k+1}{2}\biggr)\Gamma\biggl(\frac{3+k-d}{2}\biggr).
\end{eqnarray}
As $d\rightarrow 4$, this has the well-known leading term
\begin{displaymath}
I_{MB}^{IR}(d\rightarrow 4) = \frac{1}{16\pi}\biggl(\frac{M}{m}\biggr) + \ldots,
\end{displaymath}
where the dots indicate terms of higher order in $M/m$. This is clearly in 
accord with the low-energy power counting. In contrast to that, the first term 
of the expansion of the regular part $R$ obviously violates this counting 
when $d\rightarrow 4$. The infrared regularization now prescribes to drop 
$R$ from the loop contribution and substitute $I_{MB}^{IR}$ for $I_{MB}$. 
Moreover, the poles of $I_{MB}^{IR}$ in $d-4$ are also absorbed in a 
renormalization of the masses and coupling constants of the effective 
Lagrangian. Again, for a more general and comprehensive treatment of the 
IR scheme in the pion-nucleon system, the reader should consult the original 
article of Becher and Leutwyler \cite{BL}. 

\section{IR regularization for vector mesons and pions} \label{sec:PiV}
In this section, we consider another case of infrared regularization, 
first examined in \cite{BM04}. The internal baryon lines from the 
preceding section are now replaced by vector meson lines, however, 
the vector mesons do not show up as external particles in the graphs 
we consider here (an example for the treatment of such a graph can be found 
in sec.~11 of \cite{BM04}). The only external particles here are pions 
(or, in some cases, soft photons etc.), so that there are only small external 
momenta of order $O(q)$ flowing into the loop. The fundamental scalar loop 
integral is in this case
\begin{equation} \label{eq:IMV}
I_{MV}(q^2) = \int\frac{d^{d}l}{(2\pi)^{d}}\frac{i}{((q-l)^2-M_{V}^2)(l^2-M^2)} ,
\end{equation}
where $M_{V}$ is the mass of the heavy meson resonance and $q$ is some small 
external momentum (small with respect to the resonance mass $M_{V}$). This is, 
in principle, the same function as in eq.~(\ref{eq:IMB}), so the reader might 
ask why we devote this section to the examination of this case. The point is
that the extraction of the infrared singular part of $I_{MV}$ must proceed 
along different lines here. As shown in \cite{BM04}, a splitting like that 
of eq.~(\ref{eq:split}) does not amount to a separation into infrared singular 
and regular parts as in the preceding section. This can be traced back to the 
fact that the extension of the parameter integrals to infinity leads to a
singular behaviour of the loop function near $q^{2}= 0$. Of course, 
also $I_{MB}^{IR}$ has such a singularity as the external momentum squared 
goes to zero (see sec.~5.4 of \cite{BL}), but the point $p^{2}=0$ lies far 
outside the low-energy region in that case. Here, however, the point $q^{2}=0$ 
lies at the center of the low-energy region, so the infrared singular and
regular parts can not be expressed as parameter integrals from zero or one 
to infinity, respectively. For example, a 'regular' part defined as being 
proportional to the parameter integral from one to infinity would not be 
expandable around $q^{2}=0$, as it should be in order to be able to absorb 
the corresponding terms in a renormalization of the local operators in 
the effective Lagrangian. 

To circumvent this difficulty, we will take up the following simple idea 
from \cite{BM04}. We observe that $I_{MV}$ is analytic at $q^{2}=0$, the only 
singularity being the threshold branch point at $q^{2}=(M_{V}+M)^{2}$, which
is far outside the range of small $q^{2}$-values. Expanding $I_{MV}$ in
$q^{2}$, the analyticity properties in that variable are obvious. Each 
coefficient in this expansion can be split into an infrared singular and a 
regular part almost like in eq.~(\ref{eq:split}). Extracting the infrared 
singular part proportional to $d$-dependent powers of the pion mass of each 
coefficient, and resumming the series, one arrives at a well-defined 
expression for the infrared singular part of the loop integral $I_{MV}$ 
that is (by construction) expandable in the small variable $q^{2}/M_{V}^{2}$.

To start with the analysis, let us first consider the special case where 
the external momentum vanishes: $q=0$. Then we have
\begin{eqnarray*}
I_{MV}(0) &=& \int\frac{d^{d}l}{(2\pi)^{d}}\frac{i}{(l^2-M_{V}^2)(l^2-M^2)} \\ 
&=& \frac{1}{M^{2}-M_{V}^{2}}\biggl(\int\frac{d^{d}l}{(2\pi)^{d}}
\frac{i}{l^{2}-M^{2}}-\int\frac{d^{d}l}{(2\pi)^{d}}\frac{i}{l^{2}-M_{V}^{2}}\biggr).
\end{eqnarray*}
In the last step of this equation, we have already achieved the splitting 
into an infrared singular and a regular part, which is quite trivial here. 
The first term is proportional to a $d$-dependent power of $M$ 
(compare eq.~(\ref{eq:IM})), while the second part is clearly expandable 
in $M^{2}$ (because $M_{V}^{2}\gg M^{2}$). Moreover, the pion mass expansion
of the first term starts with $M^{d-2}$, in agreement with the low-energy 
power counting for $I_{MV}$: In contrast to the baryon propagator in 
sec.~\ref{sec:PiN}, the resonance propagator is counted as $O(q^{0})$ since, 
in the low-energy region, the off-shell momentum of the internal resonance 
line is far below its  mass shell. The counting for the pion propagator and
the loop measure is the same as before, and one is lead to the power counting 
result $q^{d-2}$ for $I_{MV}$. Therefore we can write
\begin{equation} \label{eq:IMVIR0}
I_{MV}^{IR}(0) = \frac{1}{M^{2}-M_{V}^{2}}\int\frac{d^{d}l}{(2\pi)^{d}}
\frac{i}{l^{2}-M^{2}}.
\end{equation}
It should be clear that there will only be corrections of $O(q^{2})$ to 
this result when we compute the regularized integral for nonvanishing $q$. 
We introduce the following variables,
\begin{equation} \label{eq:ab}
\tilde\alpha = \frac{M^{2}}{M_{V}^{2}}\qquad,\qquad \tilde\beta = \frac{q^{2}}{M_{V}^{2}},
\end{equation}
 which we assume to be small in the sense that $\tilde\alpha,\tilde\beta\ll
 1$. Just like in sec.~\ref{sec:PiN}, we use the Feynman parameter trick to write
\begin{equation}
I_{MV}(q^{2}) = -M_{V}^{d-4}\frac{\Gamma(2-\frac{d}{2})}{(4\pi)^{\frac{d}{2}}}
\int_{0}^{1}\frac{dz}{(\tilde\beta z^{2}+z(1
-\tilde\alpha-\tilde\beta)+\tilde\alpha)^{2-\frac{d}{2}}}.
\end{equation}
We rewrite this expression as follows:
\begin{displaymath}
I_{MV}(q^{2}) =
-M_{V}^{d-4}\frac{\Gamma(2-\frac{d}{2})}{(4\pi)^{\frac{d}{2}}}
\int_{0}^{1}\frac{dz}{(z(1-\tilde\alpha-\tilde\beta)
+\tilde\alpha)^{2-\frac{d}{2}}}\biggl(1+\frac{\tilde\beta z^{2}}{z(1
-\tilde\alpha-\tilde\beta)+\tilde\alpha}\biggr)^{\frac{d}{2}-2} .
\end{displaymath}
This can be expanded according to
\begin{displaymath}             
I_{MV}(q^{2}) = -\frac{M_{V}^{d-4}}{(4\pi)^{\frac{d}{2}}}
\sum_{k=0}^{\infty}\frac{\Gamma(\frac{d}{2}-1)}{k!\Gamma(\frac{d}{2}-1-k)}
\int_{0}^{1}\frac{\Gamma(2-\frac{d}{2})dz}{(z(1-\tilde\alpha-\tilde\beta)
+\tilde\alpha)^{2-\frac{d}{2}}}
\biggl(\frac{\tilde\beta z^{2}}{z(1-\tilde\alpha-\tilde\beta)+\tilde\alpha}\biggr)^{k}.
\end{displaymath}
In the next step, we extract the infrared singular part of each term in the
sum. In each parameter integral, we substitute $z=\tilde\alpha y$ and 
extend the integration range to infinity:
\begin{equation} \label{eq:Ik}
I_{k} \equiv \int_{0}^{1}dz \frac{z^{2k}}{(z(1-\tilde\alpha-\tilde\beta)
+\tilde\alpha)^{2+k-\frac{d}{2}}} \rightarrow 
\int_{0}^{\infty}dy\frac{\tilde\alpha^{\frac{d}{2}-1+k}y^{2k}}{(1+y(1
-\tilde\alpha-\tilde\beta))^{2+k-\frac{d}{2}}}.
\end{equation} 
Divergences of the parameter integral due to the extension of the upper 
limit to infinity are again handled by analytic continuation in $d$. 
It can be expressed in terms of Gamma functions:
\begin{displaymath}
\int_{0}^{\infty}dy\frac{y^{2k}}{(1+y(1-\tilde\alpha-\tilde\beta))^{2+k-\frac{d}{2}}}
 = \frac{\Gamma(2k+1)\Gamma(1-\frac{d}{2}-k)}{(1
-\tilde\alpha-\tilde\beta)^{2k+1}\Gamma(2-\frac{d}{2}+k)}.
\end{displaymath}
This is clearly expandable in the small variables $\tilde\alpha$ and 
$\tilde\beta$, so that the r.h.s. of eq.~(\ref{eq:Ik}) in fact has the 
proper form of an infrared singular contribution, being proportional 
to $d$-dependent powers of $\tilde\alpha$. Moreover, it is not difficult 
to see that the parameter integrals from $1$ to infinity are completely 
regular in the small variables. So, putting pieces together, and using 
the following identity for Gamma functions:
\begin{equation} \label{eq:GammaId}
\frac{\Gamma(\frac{d}{2}-1)}{\Gamma(\frac{d}{2}-1-k)} 
= (-1)^{k}\frac{\Gamma(2-\frac{d}{2}+k)}{\Gamma(2-\frac{d}{2})}, \qquad k\in\mathbf{N},
\end{equation}
we can sum the series of infrared singular terms and write
\begin{equation} \label{eq:IMVIR}
I_{MV}^{IR}(q^{2}) = -\frac{M_{V}^{d-4}}{(4\pi)^{\frac{d}{2}}}
(\tilde\alpha)^{\frac{d}{2}-1}\sum_{k=0}^{\infty}\frac{(-\tilde\alpha\tilde\beta)^{k}}{(1
-\tilde\alpha-\tilde\beta)^{2k+1}}\frac{\Gamma(2k+1)\Gamma(1-\frac{d}{2}-k)}{\Gamma(k+1)}.
\end{equation}
Note that the leading term in this result is of chiral order $O(q^{d-2})$, as 
predicted by the power counting scheme. All the terms we have separated off 
from $I_{MV}$ are regular in both small parameters, and only the infrared
singular 
terms remain. For $\tilde\beta=0$, the result for $I_{MV}^{IR}(0)$, 
which was already established in eq.~(\ref{eq:IMVIR0}), is reproduced. 

It is perhaps worth noting that the result of eq.~(\ref{eq:IMVIR}) can be 
obtained in a different way, which goes back to Ellis and Tang
\cite{Ellis,Tang}. 
Though they use their method in the pion-nucleon sector, a variant of it is 
also applicable here. Their prescription to obtain the soft momentum 
contribution of a loop integral is the following: Expand the propagators of
the heavy particles {\em as if\/} the loop momentum were small, and then
interchange summation and loop integration. It is claimed that this
prescription eliminates the 'hard momentum' contributions present in 
the full loop graph. If this is true, and the concept of hard vs. soft
momentum effects is a well-defined one, the result of the procedure should 
reproduce the infrared singular part of the loop integral in question. 
Let us see how this works out for our example. Following the prescription 
just described step by step, we make the following set of transformations:
\begin{eqnarray*}
I_{MV}(q^{2}) &\rightarrow& \int\frac{d^{d}l}{(2\pi)^{d}}\frac{i}{l^{2}-M^{2}}
\sum_{k=0}^{\infty}\frac{(2q\cdot l)^{k}}{(q^{2}+l^{2}-M_{V}^{2})^{k+1}} \\
&\rightarrow& \int\frac{d^{d}l}{(2\pi)^{d}}\frac{i}{l^{2}-M^{2}}
\sum_{k=0}^{\infty}\frac{(2q\cdot l)^{k}}{(q^{2}+M^{2}-M_{V}^{2})^{k+1}} \\ 
&\rightarrow& \sum_{k=0}^{\infty}\int\frac{d^{d}l}{(2\pi)^{d}}
\frac{i(2q\cdot l)^{k}}{(l^{2}-M^{2})(q^{2}+M^{2}-M_{V}^{2})^{k+1}} .\\
\end{eqnarray*}
While the first step is just the expansion of the vector meson propagator 
pole imposed by the prescription, the second step deserves a comment: There, 
we have used the same trick of partial fractions to split off some hard 
momentum contributions as in the treatment of $I_{MV}(0)$, but now this 
was performed $k+1$ times. In the last step, summation and integration 
were interchanged. Using the formula
\begin{equation} \label{eq:supertadpole}
\int\frac{d^{d}l}{(2\pi)^{d}}\frac{i(q\cdot l)^{2n}}{l^{2}-M^{2}} 
= (-q^{2}M^{2})^{n}M^{d-2}\frac{\Gamma(n+\frac{1}{2})}{
\Gamma(\frac{1}{2})}\frac{\Gamma(1-\frac{d}{2}-n)}{(4\pi)^{\frac{d}{2}}}  
\end{equation}
and the fact that the loop integrals in the series vanish if $k$ is odd, 
the result of the transformation is 
\begin{equation} \label{eq:IVMsoft}
I_{MV}^{\mathrm{soft}}(q^{2}) = \frac{M^{d-2}}{(4\pi)^{\frac{d}{2}}}
\sum_{n=0}^{\infty}\frac{(-4q^{2}M^{2})^{n}}{(q^{2}+M^{2}-M_{V}^{2})^{2n+1}}
\frac{\Gamma(n+\frac{1}{2})\Gamma(1-\frac{d}{2}-n)}{\Gamma(\frac{1}{2})}.
\end{equation}
Extracting a factor of $M_{V}^{d-4}$, and using the identity
\begin{equation} \label{eq:GammaId2}
\frac{\Gamma(2n+1)}{\Gamma(n+1)} 
= 4^{n}\frac{\Gamma(n+\frac{1}{2})}{\Gamma(\frac{1}{2})},
\end{equation}
we see by comparing to eq.~(\ref{eq:IMVIR}) that $I_{MV}^{\mathrm{soft}} =
I_{MV}^{IR}$. 

In \cite{BM04}, the result for $I_{MV}^{IR}$ was given in a different form 
(see eqs.~(7.10) and (8.4) of that reference). Of course, it is equivalent 
to the result derived above: it is shown in app.~\ref{app:closedform} that the 
two different forms just amount to a reordering of the corresponding
expansions. We will find that the form of eq.~(\ref{eq:IMVIR}) is most 
practical for our purposes, in particular, the chiral expansion can almost 
immediately be read off from that formula.

In closing this section, we note that the spin or the parity of the 
resonance obviously do not play a major role in the above considerations. 
Though we will concentrate on the case of vector mesons, most of what we 
have said would also apply for other meson resonances, 
like e.g. scalar or axial-vector mesons.
 
\section{Pion-nucleon system with explicit meson resonances} 
\label{sec:PiNV}

Now that we have collected the results for the fundamental one-loop 
integrals in the pion-nucleon and the vector meson-pion sector, we are 
prepared to extend the framework of infrared regularization once more, and 
apply it to Feynman graphs where nucleons, pions as well as vector mesons 
take part in the same loop. The simplest example where this situation occurs 
is the triangle graph consisting of one pion, one nucleon and one vector meson 
line. Due to baryon number conservation, the nucleon line must run through 
the complete diagram. We shall assume that only a small momentum $k$ (small 
in the usual sense) is transferred at the vector meson-pion vertex. Such a 
graph will typically contribute to some nucleon form factor in the region of 
small momentum transfer. With this application in mind, and with the excuse 
that it will simplify the presentation a bit, we will further specify to 
on-shell nucleons. Let $p$ and $\bar p$ be the four-momenta of the incoming 
and the outgoing nucleon. Then we have
\begin{equation} \label{eq:kinppbar}
p^{2} = m^{2}=\bar p^{2}
=(p+k)^{2} \Rightarrow k^{2}=2\bar p\cdot k=-2p\cdot k.
\end{equation} 
Using these kinematic relations, we can rewrite the fundamental 
scalar loop integral
\begin{equation} \label{eq:IMBV}
I_{MBV}(k^{2}) \equiv \int\frac{d^{d}l}{(2\pi)^{d}}
\frac{i}{((p-l)^{2}-m^{2})((k+l)^{2}-M_{V}^{2})(l^{2}-M^{2})}
\end{equation}
with the help of the usual Feynman parameter trick, as
\begin{equation} \label{eq:intlxy}
\int\frac{d^{d}l}{(2\pi)^{d}}\int_{0}^{1}\int_{0}^{1-y}
\frac{2idxdy}{(y((p-l)^{2}-m^{2})+x((k+l)^{2}-M_{V}^{2})+(1-x-y)(l^{2}-M^{2}))^{3}}.
\end{equation}
Doing the loop integration in the usual manner, we get
\begin{equation} \label{eq:intxy}
I_{MBV}(k^{2}) 
= m^{d-6}\frac{\Gamma(3-\frac{d}{2})}{(4\pi)^{\frac{d}{2}}}
\int_{0}^{1}\int_{0}^{1-y}\frac{dxdy}{(y^{2}-\alpha y+\alpha 
+ \beta(x^{2}+xy)+x(\gamma-\alpha-\beta))^{3-\frac{d}{2}}},
\end{equation}
where the definitions
\begin{displaymath}
\alpha = \frac{M^{2}}{m^{2}} \ll 1, \qquad \beta  
= \frac{k^{2}}{m^{2}} \ll 1, \qquad 
\gamma = \frac{M_{V}^{2}}{m^{2}} \sim O(1). 
\end{displaymath}
 were used. In particular, we have assumed here that the nucleon 
and the meson resonance are roughly of the same order of magnitude 
(in the real world, we have $\gamma \sim 2/3$ for the rho resonance, 
which is good enough for our purposes).\\
We should remark here that there is, of course, a second graph with 
the same topology, where the pion and the resonance line are interchanged
(see fig.~\ref{fig:triangles}). 
The expression for the corresponding scalar loop integral is 
\begin{equation} \label{eq:IMBV2}
\tilde I_{MBV}(k^{2}) \equiv 
\int\frac{d^{d}l}{(2\pi)^{d}}\frac{i}{((\bar p-l)^{2}-m^{2})((l-k)^{2}-M_{V}^{2})(l^{2}-M^{2})}.
\end{equation}
However, the reader can convince himself that, due to the on-shell 
kinematics specified in eq.~(\ref{eq:kinppbar}), this will give exactly 
the same expression as in eq.~(\ref{eq:intxy}). Thus, we can focus 
on the integral $I_{MBV}$.

In analogy to sec.~\ref{sec:PiV}, it will be instructive to begin with 
the special case where $k=0$. Since we must split off the terms where 
only propagators of heavy particles occur, we can obviously apply the 
same partial fraction method that led to eq.~(\ref{eq:IMVIR0}). 
The remaining pion-nucleon integral can be dealt with as 
in sec.~\ref{sec:PiN}. This gives
\begin{equation} \label{eq:IMBVIR0}
I_{MBV}^{IR}(0) = \frac{I_{MB}^{IR}(m^{2})}{M^{2}-M_{V}^{2}}.
\end{equation}
Having the standard method described of sec.~\ref{sec:PiN} in mind, 
we note that this equals
\begin{eqnarray*}
I_{MBV}^{IR}(0) &=& m^{d-6}\frac{\Gamma(3-\frac{d}{2})}{(4\pi)^{\frac{d}{2}}}
\int_{0}^{\infty}\int_{0}^{\infty}\frac{dxdy}{(y^2-\alpha y+\alpha
+x(\gamma-\alpha))^{3-\frac{d}{2}}} \\
 &=& \frac{m^{d-6}}{(\gamma-\alpha)}
\frac{\Gamma(2-\frac{d}{2})}{(4\pi)^{\frac{d}{2}}}
\int_{0}^{\infty}\frac{dy}{(y^{2}-\alpha y+\alpha)^{2-\frac{d}{2}}} 
\end{eqnarray*}
(compare the last line with the l.h.s of eq.~(\ref{eq:IMBIR})). Once again,
the possible divergence for large $d$ at $x\rightarrow\infty$ was regularized
by analytic continuation from small $d$ as before. It is reassuring to see
that eq.~(\ref{eq:IMBVIR0}) is reproduced in this way. 

\FIGURE[t]{\includegraphics[width=7cm]{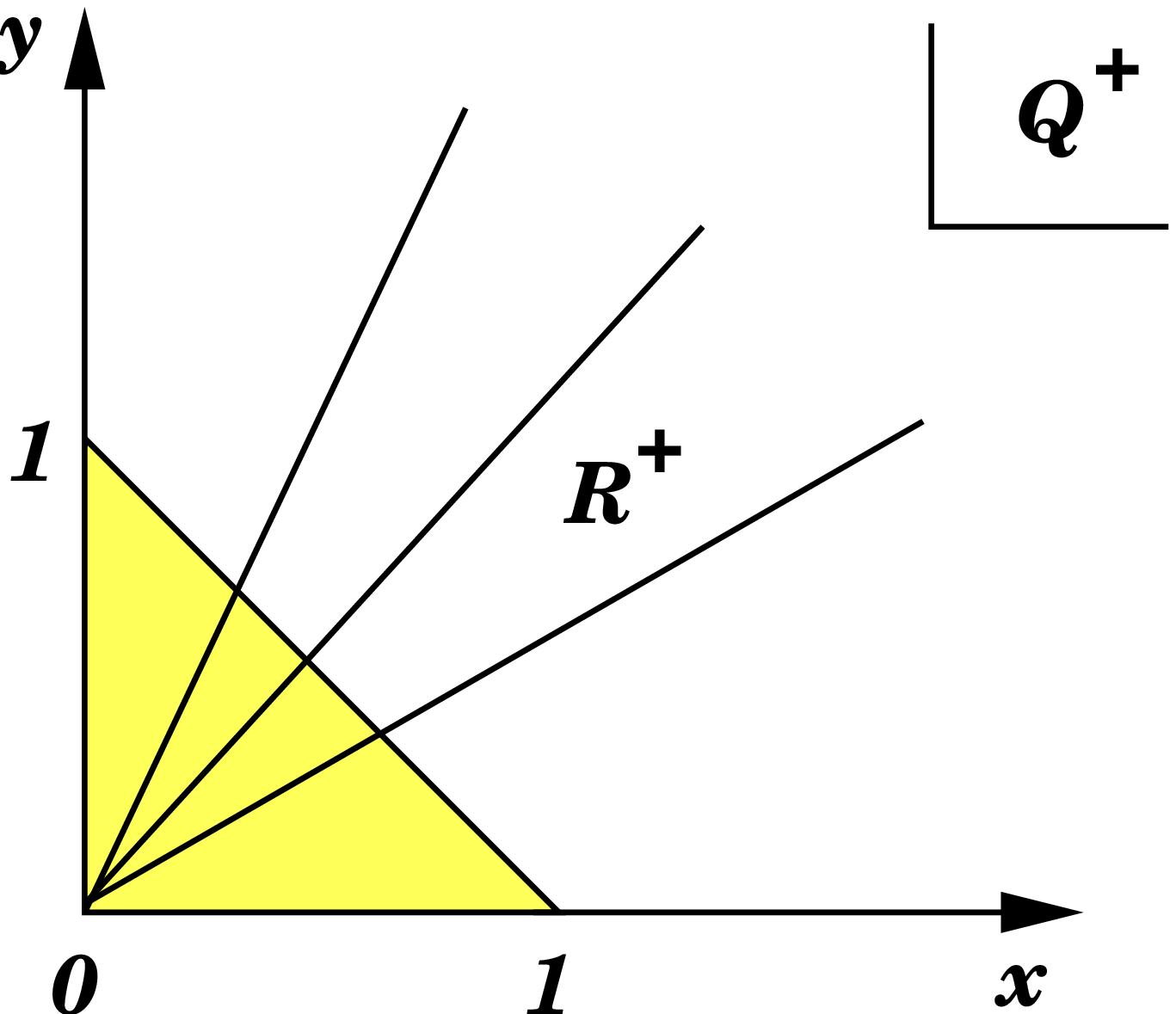}
\caption{Illustration of the integration ranges for the infrared singular 
and the regular part of $I_{MBV}$. The shaded triangle is the region
integrated over in eq.~(\ref{eq:intxy}). The three rays indicate 
the extension of the connections from the soft point at $(0,0)$ 
to the hard line from $(0,1)$ to $(1,0)$.}
\label{fig:rays}}

The extension of both parameter integrations to infinity is the natural 
generalization of the prescription used in sec.~\ref{sec:PiN}. This can be 
seen as follows. In eq.~(\ref{eq:intlxy}), we chose our Feynman parameters 
such that the pion propagator contributes with the weight one at $x=y=0$, 
while the vector meson and the nucleon propagator have their maximum 
weight at $(x,y) = (1,0)$ and $(0,1)$, respectively. Consequently, the 
infrared singularity is located in parameter space at the point $(x,y) =
(0,0)$. Indeed, a look at eq.~(\ref{eq:intxy}) confirms that the integrand 
in that expression is approximately $\alpha^{\frac{d}{2}-3}$ in the region 
where  $x\sim y\sim 0$, thus producing infrared singular terms there. Far from 
this region, the integrand is expandable in $\alpha$, as we will show later. 
To be more specific, imagine the positive quarter of the parameter plane, 
i.e. $Q_{+}\equiv \lbrace(x,y):x\geq 0, y\geq 0\rbrace$, split in two parts, 
the first being the triangle one integrates over in the full loop integral 
(see eq.~(\ref{eq:intxy})) and the second part its complement in $Q_{+}$, 
named $R_{+}$ (see fig.~\ref{fig:rays} for an illustration). In the latter 
parameter region, no infrared singularities are located, hence, the
integration over this region should yield a proper regular part (there is 
a qualification here, see below). Geometrically, one might imagine that all 
lines from the 'soft point' $(0,0)$ (the location of the infrared singularity 
in parameter space) to the 'hard line' from $(0,1)$ to $(1,0)$ are extended to 
infinity to get the infrared singular part of the loop integral. In a more 
general setting, there may be soft or hard points, lines, surfaces etc., 
depending on the number of soft and hard pole structures in the integral 
under consideration. Extending all connecting lines from the soft point, 
line etc. to the hard point (line$\ldots$) to infinity, one always achieves 
a splitting in the original parameter region and a region where no infrared 
singularities in parameter space are present, analogous to $R_{+}$. It is
possible to show that this geometric picture leads to the same results as 
the prescription given in sec.~6 of \cite{BL}. This picture might serve 
as a guide when trying to split arbitrary one-loop graphs into infrared 
singular and regular parts, however, one should always convince oneself that 
both those parts have the correct properties, and that their sum equals the 
original integral. One has to be a bit careful here, as the following
qualification shows. As explained in sec.~\ref{sec:PiV}, extending the 
integration range to infinity might lead to unphysical singularities in 
variables in which the original integral is analytic (at least in the 
low-energy region). Those singularities will be harmless if they are located 
far from the low-energy region, as e.g. the singularity at $s=0$ of the 
infrared singular parts in the pion-nucleon sector, but in other cases, 
they can be disturbing. Following the method of \cite{BM04}, we avoided 
this problem in sec.~\ref{sec:PiV} by expanding the original loop integral 
in the small variable $\tilde\beta$ beforehand. Only then could the 
integration range be extended to infinity, in each coefficient of the 
expansion. We must expect that a similar phenomenon will occur in the present
case.

To exclude this from the start, we expand $I_{MBV}$ in analogy to 
eq.~(\ref{eq:IMVIR}):
\begin{eqnarray*}
I_{MBV}(k^{2}) &=& \frac{m^{d-6}}{(4\pi)^{\frac{d}{2}}}
\sum_{j=0}^{\infty}\frac{\Gamma(\frac{d}{2}-2)}{\Gamma(\frac{d}{2}-2-j)}
\int_{0}^{1}\int_{0}^{1-y}\frac{\Gamma(3-\frac{d}{2})dxdy(\beta x(x+y))^{j}}{j!(y^2
-\alpha y+\alpha+x(\gamma-\alpha-\beta))^{3-\frac{d}{2}+j}} \\
&=& \frac{m^{d-6}}{(4\pi)^{\frac{d}{2}}}
\sum_{j=0}^{\infty}\frac{\Gamma(\frac{d}{2}-2)}{\Gamma(\frac{d}{2}-2-j)}
\int_{0}^{1}\int_{0}^{1-y}\frac{\Gamma(3-\frac{d}{2})dxdy(\sum_{l=0}^{j}{j\choose l}
\beta^{j}x^{l+j}y^{j-l})}{j!(y^2-\alpha y
+\alpha+x(\gamma-\alpha-\beta))^{3-\frac{d}{2}+j}}. 
\end{eqnarray*}
There is still some $\beta$-dependence in the denominator, but in 
that combination, it will turn out to be harmless. We extend the 
parameter integrations to $Q_{+}$ and define
\begin{displaymath}
I_{MBV}^{IR}(k^{2}) =
\frac{m^{d-6}}{(4\pi)^{\frac{d}{2}}}
\sum_{j=0}^{\infty}\frac{\Gamma(\frac{d}{2}-2)}{
\Gamma(\frac{d}{2}-2-j)}\int_{0}^{\infty}\int_{0}^{\infty}
\frac{\Gamma(3-\frac{d}{2})dxdy(\sum_{l=0}^{j}{j\choose l}
\beta^{j}x^{l+j}y^{j-l})}{j!(y^2-\alpha y+\alpha+x(\gamma
-\alpha-\beta))^{3-\frac{d}{2}+j}}.  
\end{displaymath}
Since the denominator is always positive for $0<\alpha\ll 1, |\beta|\ll 1$, 
the $x$-integration can readily be done using the formula
\begin{equation} \label{eq:xparint}
\int_{0}^{\infty}dx\frac{x^{n}}{(a+bx)^{D}} 
= \frac{a^{n+1-D}}{b^{n+1}}\frac{\Gamma(n+1)\Gamma(D-(n+1))}{\Gamma(D)},
\end{equation}
together with eq.~(\ref{eq:GammaId}). This gives 
\begin{eqnarray*}
I_{MBV}^{IR}(k^{2}) 
&=& \frac{m^{d-6}}{(4\pi)^{\frac{d}{2}}}\sum_{j=0}^{\infty}\sum_{l=0}^{j}
\frac{\Gamma(j+l+1)\Gamma(2-\frac{d}{2}-l)}{\Gamma(j-l+1)
\Gamma(l+1)}\frac{(-\beta)^{j}}{(\gamma-\alpha-\beta)^{j+l+1}}\times \\
& & \times\int_{0}^{\infty}\frac{y^{j-l}dy}{(y^{2}-\alpha y+\alpha)^{2-\frac{d}{2}-l}}.
\end{eqnarray*}
In the next step, we can write down the chiral expansion of 
the $y$-integral, using the same method as in sec.~\ref{sec:PiN}. 
The generalized formula is
\begin{equation} \label{eq:yparint}
\int_{0}^{\infty}dy\frac{y^{n}}{(y^{2}-\alpha y+\alpha)^{D}} 
= \sqrt{\alpha}^{n+1-2D}\sum_{k=0}^{\infty}\frac{\sqrt{\alpha}^{k}}{k!}
\frac{\Gamma(\frac{n+k+1}{2})\Gamma(\frac{2D+k-(n+1)}{2})}{2\Gamma(D)}.
\end{equation} 
Obviously, the chiral expansion of this integral can straightforwardly 
be read off from the series on the r.h.s. Inserting this result, we get
\begin{equation} \label{eq:IMBVIR}
I_{MBV}^{IR}(k^{2}) = \frac{m^{d-6}}{(4\pi)^{\frac{d}{2}}}
\sum_{j,k=0}^{\infty}\sum_{l=0}^{j}\frac{(-\beta)^{j}
\sqrt{\alpha}^{d-3+j+l+k}}{(\gamma-\alpha-\beta)^{j+l+1}}
\frac{\Gamma(j+l+1)\Gamma(\frac{j+k-l+1}{2})
\Gamma(\frac{3-d-l-j+k}{2})}{2\Gamma(j-l+1)\Gamma(k+1)\Gamma(l+1)}.
\end{equation}
The only expressions we have not expanded in the small variables are the 
factors of $(\gamma-\alpha-\beta)$ in the denominator, but this can of course 
be done: the corresponding geometric series is absolutely convergent due to
the assumption that $\gamma\sim O(1)$. 

To complete the proof that eq.~(\ref{eq:IMBVIR}) is the correct infrared 
singular part of $I_{MBV}$, we have to show that all the terms we dropped 
in the extraction procedure described above are regular in $\alpha$. 
Those terms are proportional to parameter integrals over the region 
$R_{+}$, of the general type
\begin{eqnarray*}
R_{j} &=& \int_{R_{+}}\frac{dxdy(x(x+y))^{j}}{(y^{2}
-\alpha y+\alpha+x(\gamma-\alpha-\beta))^{3-\frac{d}{2}+j}} \\
&=& \int_{z=1}^{\infty}\int_{x=0}^{z}\frac{dxdz(xz)^{j}}{((z-x)^2
-\alpha(z-1)+x(\gamma-\beta))^{3-\frac{d}{2}+j}}.
\end{eqnarray*}
Here we have traded the variable $y$ for $z\equiv x+y$. One finds that the function
\begin{displaymath}
f(z,x) = \frac{(z-x)^{2}+x(\gamma-\beta)}{z-1}
\end{displaymath}
has the property 
\begin{displaymath}
f(z,x) \geq \mathrm{min}\lbrace 4,(\gamma-\beta)\rbrace
\end{displaymath}
in $R_{+}$, provided that the parameters $\beta$ and $\gamma$ are in their 
typical low-energy ranges. This is already sufficient to ensure that the 
integrand of $R_{j}$ can safely be expanded in $(z-1)\alpha$, and that
integration and summation of the corresponding series can be interchanged. 
This proves the regularity of the integrals $R_{j}$ in $\alpha$. 

It is also possible to show that the result for $I_{MBV}^{IR}$ can be 
obtained in a way that is closely analogous to the prescription of 
Ellis and Tang (see the end of sec.~\ref{sec:PiV}). The proof can be 
found in app.~\ref{app:TangIR}.

\section{Application: Axial form factor of the nucleon} 
\label{sec:AxFF}

A typical example for an application where the triangle graph 
treated in sec.~\ref{sec:PiNV} shows up is a contribution to some form 
factor of the nucleon at low momentum transfer. As a specific example, 
we consider the nucleon form factor of the isovector axial-vector current 
in a theory with an explicit rho resonance field. A representation for 
this form factor, using the infrared regularization scheme in the pion-nucleon 
sector, has been given by Schweizer \cite{Schweizer}. In that framework, 
the contributions due to the various baryon or meson resonances are contained 
in the low-energy coefficients (LECs) of the effective pion-nucleon
Lagrangian, or, more correctly: The contributions from tree-level resonance 
exchange can be described as an infinite sum of contact terms derived from 
the pion-nucleon Lagrangian. The inclusion of explicit resonance fields 
therefore amounts to a resummation of higher order terms, which is often 
advantageous (as discussed in detail in ref.~\cite{Kubis:2000zd}).
Moreover, a theory with explicit resonance fields can serve to achieve 
an understanding of the numerical values of the LECs, relying on the 
assumption that the lowest-lying resonances give the dominant contributions 
to those coefficients. This is usually called the principle of resonance 
saturation, and has been very successfull in ChPT, see
e.g. \cite{RoleofRes,AspectsPiN}. 
Assuming this principle to be valid, the pion-nucleon LECs can be expressed 
through the masses of the resonances and the couplings of the resonances 
to the nucleons and pions. Such relations are most useful, of course, 
if the resonance masses and couplings are sufficiently well known. 
\FIGURE[h]{\includegraphics[width=11cm]{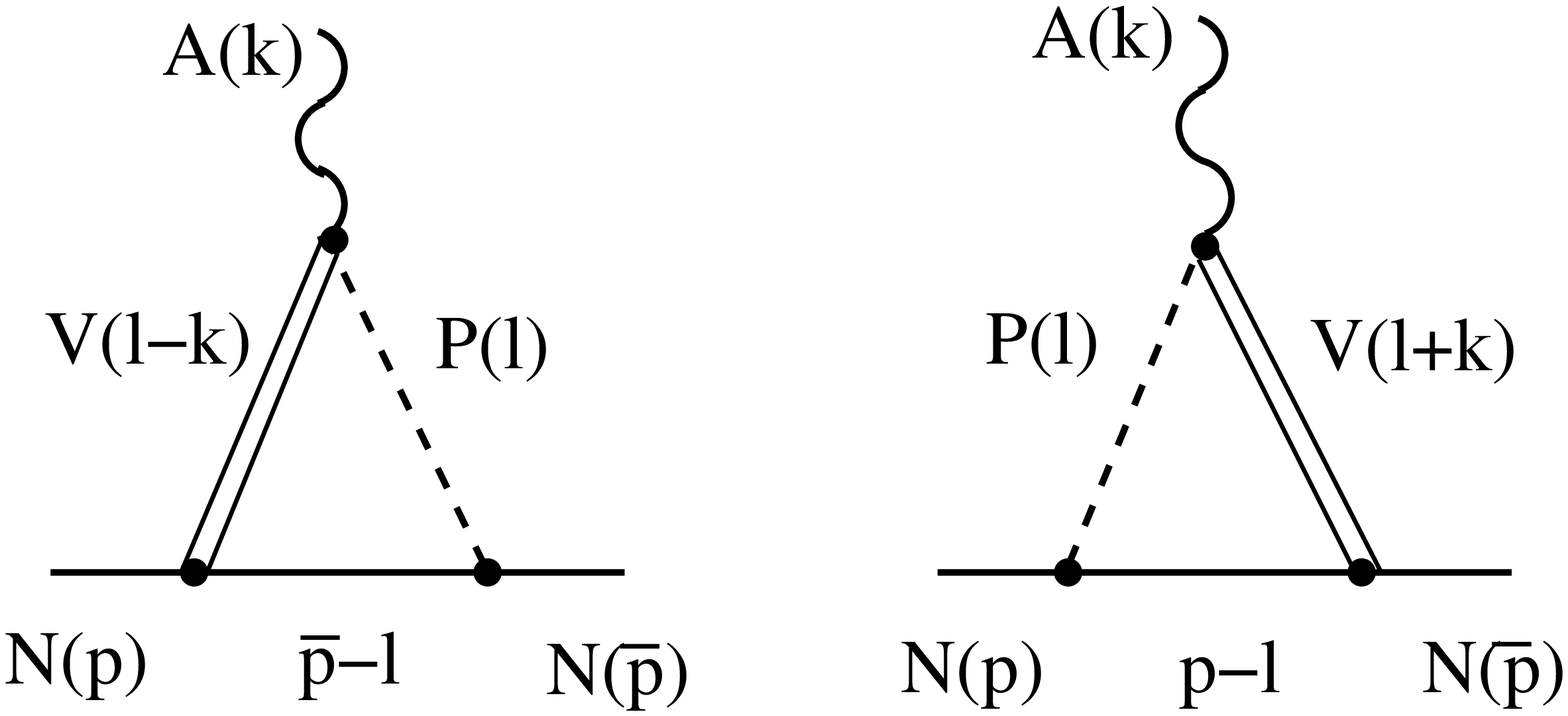}
\caption{Triangle graphs contributing to the axial form factor of the
nucleon. The dashed line represents the pion, while the double line 
stands for the vector meson. The letter $A$ indicates the external axial 
source, $N$ denotes the nucleon under consideration.}
\label{fig:triangles}}
In this section, we will compute the contribution to the axial form factor 
of the nucleon that is described by the two triangle graphs of 
fig.~(\ref{fig:triangles}), with one nucleon, one pion and one resonance line, 
in the framework of the extended infrared regularization scheme developed in 
the preceding sections. First, we must set up the necessary formalism and
collect the various terms from the effective Lagrangian we need for the
computation.

By Lorentz invariance, the matrix element of the axial current
\begin{displaymath}
A_{\mu}^{i}(x) \equiv \bar q(x)\gamma_{\mu}\gamma_{5}\frac{\tau^{i}}{2}q(x)
\end{displaymath}
between one-nucleon states can be parametrized as 
\begin{equation} \label{eq:AxMatEl}
\langle N'(\bar p) | A_{\mu}^{i}(x) | N(p)\rangle 
= \bar u'(\bar p)\biggl(G_{A}(t)\gamma_{\mu}
+G_{P}(t)\frac{k_{\mu}}{2m}+G_{T}(t)
\frac{\sigma_{\mu\nu}k^{\nu}}{2m}\biggr)\gamma_{5}\frac{\tau^{i}}{2}u(p)e^{ikx}.
\end{equation}
In the above expressions, $q$ is the quark field spinor, $q^{T}=(u,d)$, $N'$
is the outgoing nucleon with momentum $\bar p$, $N$ labels the incoming
nucleon with momentum $p$, and $k=\bar p-p$, $t\equiv k^{2}$. The 
symbols $\tau^{i}$ denote the usual Pauli matrices. 
Finally, $\bar u'$ and $u$ are the Dirac spinors associated with the 
outgoing and incoming nucleon, respectively. Assuming perfect isospin 
symmetry and charge conjugation invariance, as we will do here, 
leads to $G_{T}\equiv 0$. The relation of $G_{A}$ and $G_{P}$ to the 
quantities $F_{1,2}$ used in \cite{Schweizer} is
$F_{1}(t) = G_{A}(t), 2mF_{2}(t) = -G_{P}(t)$.
For later reference, we give the representation of $G_{A}$ up to order 
$q^{3}$ that can be found in \cite{Schweizer}:
\begin{eqnarray} \label{eq:GaSchweizer}
G_{A}(t) &=& g_{A}+4\bar d_{16}M^{2}+d_{22}t-
\frac{g_{A}^{3}M^{2}}{16\pi^{2}F^{2}} \nonumber \\ 
&+& \frac{g_{A}M^{3}}{24\pi mF^{2}}\biggl(3+3g_{A}^{2}-4c_{3}m+8c_{4}m\biggr) + O(q^{4}). 
\end{eqnarray}
Here $F$ and $g_{A}$ denote the pion decay constant and the nucleon axial charge in
the chiral limit, respectively, while the coefficients $c_{i},d_{i}$ are LECs
showing up in the pion-nucleon effective Lagrangian at order two and three,
respectively. For a precise definition of the underlying Lagrangian see
ref.~\cite{Fettes:1998ud}.

We now turn back to the calculation of the axial form factor in the presence
of vector mesons. We write down the relevant terms in the effective Lagrangian
and give the necessary rules for the vertices and propagators required for the 
calculation. First, the lowest order chiral Lagrangian for 
the pion-nucleon interaction reads 
\begin{equation} \label{eq:L1N}
\mathcal{L}^{(1)}_{N} 
= \bar\psi(i\slashed{D}-m)\psi + \frac{g_{A}}{2}\bar\psi \slashed{u}\gamma_{5}\psi.
\end{equation}
Here, $\psi$ is the nucleon spinor, $\psi^{T}=(p,n)$, and the matrix $u_{\mu}$ 
collects the pion fields $\pi^{a}$ via
\begin{eqnarray*}
u &=& \exp{\biggl(\frac{i\vec\tau\cdot\vec\pi}{2F}\biggr)}, \\
u_{\mu} &=& i\lbrace u^{\dagger},\partial_{\mu}u\rbrace 
+ u^{\dagger}r_{\mu}u-ul_{\mu}u^{\dagger}, \\
r_{\mu} &=& v_{\mu}+a_{\mu},\quad l_{\mu} = v_{\mu}-a_{\mu}.
\end{eqnarray*}
In the last line, we have introduced external isovector vector 
and axial-vector sources, $v_{\mu}$ and $a_{\mu}$,
\begin{displaymath}
v_{\mu} = v_{\mu}^{i}\frac{\tau^{i}}{2}, \quad a_{\mu} = a_{\mu}^{i}\frac{\tau^{i}}{2}.
\end{displaymath}
 The covariant derivative $D_{\mu}$ in eq.~(\ref{eq:L1N}) is defined as
\begin{eqnarray*}
D_{\mu} &=& \partial_{\mu}+\Gamma_{\mu}, \\
\Gamma_{\mu} &=& \frac{1}{2}[u^{\dagger},\partial_{\mu}u]
-\frac{i}{2}u^{\dagger}r_{\mu}u-\frac{i}{2}ul_{\mu}u^{\dagger}.
\end{eqnarray*}
>From eq.~(\ref{eq:L1N}), one derives the $\bar N N\pi$ vertex rule
\begin{displaymath}
\frac{g_{A}}{2F}\slashed{q}\gamma_{5}\tau^{a}
\end{displaymath}
for an outgoing pion of momentum $q$. 

Now we turn to the effective Lagrangians involving the vector meson fields. 
We choose a representation in terms of an antisymmetric tensor field 
$W_{\mu\nu}$ \cite{RoleofRes,Ecker,BM04}. The corresponding free Lagrangian is 
\begin{displaymath}
\mathcal{L}_{W}^{\mathrm{kin}} 
= -\frac{1}{2}\langle D^{\mu}W_{\mu\nu}D_{\rho}W^{\rho\nu}\rangle 
+ \frac{1}{4}M_{V}^{2}\langle W_{\mu\nu}W^{\mu\nu}\rangle,
\end{displaymath}
where 
\begin{displaymath}
W_{\mu\nu} = \frac{1}{\sqrt{2}}W_{\mu\nu}^{i}\tau^{i} 
= \left( \begin{array}{cc} \frac{\rho^{0}}{\sqrt{2}}\quad & \rho^{+} \\ 
\rho^{-}\quad & -\frac{\rho^{0}}{\sqrt{2}} \\ \end{array} \right)_{\mu\nu}.
\end{displaymath}
The brackets $\langle\ldots\rangle$ denote the trace in isospin space. 
{}From $\mathcal{L}_{W}^{\mathrm{kin}}$, one derives the tensor field propagator 
in momentum space,
\begin{displaymath}
T_{\mu\nu,\rho\sigma}^{ij}(k) = \frac{i\delta^{ij}}{M_{V}^{2}}
\frac{g_{\mu\rho}g_{\nu\sigma}(M_{V}^{2}-k^{2})+g_{\mu\rho}
k_{\nu}k_{\sigma}-g_{\mu\sigma}k_{\nu}k_{\rho}-(\mu 
\leftrightarrow \nu)}{M_{V}^{2}-k^{2}}.
\end{displaymath}
At lowest chiral order, the interaction of the rho meson with 
the pions is given by \cite{RoleofRes}
\begin{equation} \label{eq:LWInt}
\mathcal{L}_{W}^{\mathrm{int}} 
= \frac{F_{V}}{2\sqrt{2}}\langle F^{+}_{\mu\nu}W^{\mu\nu}\rangle 
+ \frac{iG_{V}}{2\sqrt{2}}\langle[u_{\mu},u_{\nu}]W^{\mu\nu}\rangle.
\end{equation}
In the first term we have used the definitions
\begin{eqnarray*}
F^{\pm}_{\mu\nu} &=& uF_{\mu\nu}^{L}u^{\dagger}\pm u^{\dagger}F_{\mu\nu}^{R}u, \\
F^{L}_{\mu\nu} &=& \partial_{\mu}l_{\nu}-\partial_{\nu}l_{\mu}-i[l_{\mu},l_{\nu}], \\
F^{R}_{\mu\nu} &=& \partial_{\mu}r_{\nu}-\partial_{\nu}r_{\mu}-i[r_{\mu},r_{\nu}].
\end{eqnarray*}
The external sources $r_{\mu},l_{\mu}$ are counted as $O(q)$, so that
$F^{\pm}_{\mu\nu}$ is of chiral order $O(q^{2})$. Also, $u_{\mu}$ is of
$O(q)$. Therefore, $\mathcal{L}_{W}^{\mathrm{int}}$ leads to vertices of 
chiral order $O(q^{2})$. Using the method of external sources to derive 
Greens functions from the generating functional, we must extract the 
amplitudes linear in the source $a_{\mu}$ to compute the matrix element 
of eq.~(\ref{eq:AxMatEl}). For the triangle graphs considered here, we 
need the vertex that connects the vector meson and the pion with the 
external axial source. From eq.~(\ref{eq:LWInt}), we find the 
corresponding vertex rule
\begin{displaymath}
\epsilon^{iac}\biggl(\frac{G_{V}}{F}(q_{\mu}g_{\nu\tau}-q_{\nu}g_{\mu\tau})-\frac{F_{V}}{2F}(k_{\mu}g_{\nu\tau}-k_{\nu}g_{\mu\tau})\biggr).
\end{displaymath}
Here $i,a,c$ are the isospin indices associated with the axial source
$a_{\tau}$, the pion and the vector meson field $W_{\mu\nu}$, respectively, 
$q$ is the four-momentum of the outgoing pion. Since $k$ and $q$ are counted 
as small momenta, the chiral order of this vertex rule is in accord with the 
power counting for the interaction Lagrangian. 
 
The restrictions of chiral symmetry are not that strong for the interaction 
of the vector mesons with the nucleons: Here, there are terms of chiral 
order $O(q^{0})$. The leading terms of the interaction Lagrangian have been 
given in ref.~\cite{BM95}. For the $SU(2)$ case we consider here, the 
relevant terms are
\begin{eqnarray} \label{eq:LWN}
\mathcal{L}_{NW} &=& R_{V}\bar\psi\sigma_{\mu\nu}W^{\mu\nu}\psi 
+ S_{V}\bar\psi\gamma_{\mu}D_{\nu}W^{\mu\nu}\psi \nonumber \\
& & +T_{V}\bar\psi\gamma_{\mu}D_{\lambda}W^{\mu\nu}D^{\lambda}D_{\nu}\psi 
+ U_{V}\bar\psi\sigma_{\lambda\nu}W^{\mu\nu}D^{\lambda}D_{\mu}\psi.
\end{eqnarray}
In the notation of \cite{BM95}, we have $R_{V}=R_{D}+R_{F},S_{V}=S_{D}+S_{F}$, 
etc. The definition of $\sigma_{\mu\nu}$ is standard,
$\sigma_{\mu\nu} = i[\gamma_{\mu},\gamma_{\nu}]/2.$
It turns out that only the piece proportional to $G_{V}$ from
eq.~(\ref{eq:LWInt}) 
and the piece proportional to $R_{V}$ from eq.~(\ref{eq:LWN}) contribute 
at lowest order to the diagrams computed here, the chiral expansion of 
which starts at $O(q^{3})$ (given that a scheme like infrared regularization
is used that preserves the power counting rules). To keep the presentation 
short, we show only the contribution from those terms and therefore neglect 
some higher order contributions. However, this will be sufficient to compare
our results to the representation up to $O(q^{3})$ given by Schweizer
\cite{Schweizer}.

The evaluation of the first graph gives (see fig.~(\ref{fig:triangles}))
\begin{displaymath}
I_{1} = \int\frac{d^{d}l}{(2\pi)^{d}}\biggl(\frac{G_{V}}{F}(l_{\mu}g_{\nu\tau}
-l_{\nu}g_{\mu\tau})\epsilon^{iac}\biggr)\frac{i}{l^{2}-M^{2}}
\biggl(-\frac{g_{A}}{2F}\slashed{l}\gamma_{5}\tau^{a}\biggr)\frac{iT^{\mu\nu,\rho\sigma}_{cd}(l-k)}
{\slashed{\bar p}-\slashed{l}-m}i\sigma_{\rho\sigma}
\tau^{d}\frac{R_{V}}{\sqrt{2}}, 
\end{displaymath}
and the second one gives
\begin{displaymath}
I_{2} = \int\frac{d^{d}l}{(2\pi)^{d}}\frac{R_{V}}{\sqrt{2}}
i\sigma_{\mu\nu}\tau^{c}
\frac{iT^{\mu\nu,\rho\sigma}_{cd}(l+k)}{\slashed{p}-\slashed{l}-m}\biggl(\frac{g_{A}}{2F}\slashed{l}
\gamma_{5}\tau^{a}\biggr)\frac{i}{l^{2}-M^{2}}\biggl(\frac{G_{V}}{F}(l_{\rho}
g_{\sigma\tau}-l_{\sigma}g_{\rho\tau})\epsilon^{ida}\biggr).
\end{displaymath}
We have left out the Dirac spinors $\bar u,u$ here. Since we consider on-shell 
nucleons, we can use the Dirac equation to simplify the numerators of the integrals,
$\bar u(\bar p)\slashed{\bar p}=\bar u(\bar p)m$, $\slashed{p}u(p) = mu(p)$.
We will make some remarks on the computation of $I_{1}$ (the computation of
$I_{2}$ can be done analogously). In a first step, we reduce the full loop 
integral to a linear combination of scalar loop integrals, which have been 
treated in detail in the preceding sections. Scalar loop integrals without 
a pion propagator denominator are dropped using infrared regularization, 
since they are pure regular parts. Therefore we can replace 
$l^{2}\rightarrow M^{2}$ everywhere in the numerator.

With the abbreviation 
\begin{displaymath}
g_{1} = \frac{2\sqrt{2}g_{A}G_{V}R_{V}}{M_{V}^{2}F^{2}}
\end{displaymath}
we get
\begin{eqnarray*}
I_{1} &=& g_{1}\tau^{i}\biggl(\int\frac{d^{d}l}{(2\pi)^{d}}
\frac{2im\slashed{l}((M^{2}-k\cdot l)(l^{\rho}-k^{\rho})\sigma_{\rho\tau}
+\frac{i}{2}(k_{\tau}-l_{\tau})(\slashed{l}\slashed{k}-\slashed{k}\slashed{l}))
\gamma_{5}}{((\bar p-l)^{2}-m^{2})((l-k)^{2}-M_{V}^{2})(l^{2}-M^{2})} \\
&+& \int\frac{d^{d}l}{(2\pi)^{d}}\frac{i((M^{2}-k\cdot l)(l^{\rho}-k^{\rho})
\sigma_{\rho\tau}+\frac{i}{2}(k_{\tau}-l_{\tau})(\slashed{l}\slashed{k}
-\slashed{k}\slashed{l}))\gamma_{5}}{((l-k)^{2}-M_{V}^{2})(l^{2}-M^{2})} \\
&-& \int\frac{d^{d}l}{(2\pi)^{d}}\frac{2im\slashed{l}l^{\rho}
\sigma_{\rho\tau}\gamma_{5}}{((\bar p-l)^{2}-m^{2})(l^{2}-M^{2})}\biggr).
\end{eqnarray*}
For completeness, we shall give the relevant loop integrals with tensor 
structures in app.~\ref{app:LoopInt}. Using the coefficient functions defined 
there, the result for the first integral $I_{1}$ can be written as
\begin{equation} \label{eq:I1decomp}
I_{1} = -ig_{1}\tau^{i}(\gamma_{\tau}I_{1}^{(\gamma)}+k_{\tau}I_{1}^{(k)}
+\bar p_{\tau}I_{1}^{(p)})\gamma_{5},
\end{equation}
where the coefficients read
\begin{eqnarray*}
I_{1}^{(\gamma)} &=& 2m(M_{V}^{2}-k^{2})C_{1}+mk^{2}(I_{MBV}^{A}
+I_{MBV}^{B})(M^{2}+M_{V}^{2}-k^{2})\\
& & -mM^{2}(M^{2}+M_{V}^{2}-k^{2})I_{MBV} + mM^{2}I_{MB}, \\
I_{1}^{(k)}     
&=&
2m^{2}(M_{V}^{2}-k^{2})(3C_{2}+C_{3}+4C_{4})+4m^{2}M^{2}(I_{MBV}^{A}+I_{MBV}^{B})
\\ 
& & -2m^{2}(M_{V}^{2}-k^{2})(3I_{MBV}^{A}+I_{MBV}^{B}) 
+2m^{2}(I_{MBV}^{A}-I_{MBV}^{B})(M^{2}+M_{V}^{2}-k^{2}) \\ 
& & -4m^{2}M^{2}I_{MBV} + t_{MV}^{(1)}+(M_{V}^{2}-2k^{2})I_{MV}^{(1)}+(k^{2}-M_{V}^{2})I_{MV} \\
& & +(M^{2}+M_{V}^{2}-k^{2})(I_{MV}-I^{(1)}_{MV}),\\
I_{1}^{(p)}  &=&
2m^{2}(M_{V}^{2}-k^{2})(3C_{2}-C_{3}-2C_{4})+4m^{2}M^{2}(I_{MBV}^{A}
-I_{MBV}^{B}) \\ & & -2m^{2}(I_{MBV}^{A}-I_{MBV}^{B})(M^{2}+M_{V}^{2}-k^{2})
+2t^{(0)}_{MV}-2m^{2}I_{MB}^{(1)} \\ 
& & +(M^{2}+M_{V}^{2}-k^{2})(I_{MV}^{(1)}-I_{MV})-I_{M}.
\end{eqnarray*}
What concerns the evaluation of $I_{2}$, we note that it is given by
\begin{equation} \label{eq:I2decomp}
I_{2} = -ig_{1}\tau^{i}(\gamma_{\tau}I_{2}^{(\gamma)}
+k_{\tau}I_{2}^{(k)}+p_{\tau}I_{2}^{(p)})\gamma_{5},
\end{equation}
with
\begin{displaymath}
I_{2}^{(\gamma)}=I_{1}^{(\gamma)}, \quad I_{2}^{(k)} = I_{1}^{(k)}, 
\quad I_{2}^{(p)}=-I_{1}^{(p)}.
\end{displaymath}
The sum of both graphs therefore gives
\begin{displaymath}
I_{1+2} = I_{1}+I_{2} = -ig_{1}\tau^{i}(\gamma_{\tau}I_{1+2}^{(\gamma)}
+k_{\tau}I_{1+2}^{(k)})\gamma_{5},
\end{displaymath}
with
\begin{displaymath}
I_{1+2}^{(\gamma)} = 2I_{1}^{(\gamma)}, \quad I_{1+2}^{(k)} = 2I_{1}^{(k)}+I_{1}^{(p)}.
\end{displaymath}
In the sum of the two graphs, the contribution proportional to $(\bar
p+p)_{\tau}$ cancels, as was to be expected on general grounds (see the 
remarks following eq.~(\ref{eq:AxMatEl})). We are now in a position to 
display the decomposition of the graphs as a linear combination of the scalar 
loop integrals worked out in the preceding sections:
\begin{eqnarray}
I_{1+2}^{(\gamma)} 
&=& c_{MBV}^{(\gamma)}I_{MBV}^{IR}+c_{MB}^{(\gamma)}I_{MB}^{IR}
+c_{MV}^{(\gamma)}I_{MV}^{IR}, \label{eq:I12gdecomp} \\
I_{1+2}^{(k)} 
&=& c_{MBV}^{(k)}I_{MBV}^{IR}+c_{MB}^{(k)}I_{MB}^{IR}+c_{MV}^{(k)}I_{MV}^{IR}
+c_{M}^{(k)}I_{M}. \label{eq:I12kdecomp}
\end{eqnarray}  
The expressions for the coefficients $c^{(\gamma)}$ read
\begin{eqnarray*}
c_{MBV}^{(\gamma)} 
&=& \frac{4m}{(d-2)k^{2}(k^{2}-4m^{2})}\biggl[m^{2}M_{V}^{6}
+(k^{2}((d-5)m^{2}+M^{2})-2m^{2}M^{2})M_{V}^{4} \\ 
&-& ((M^{2}+m^{2}(2d-7))k^{4}-2(d-2)m^{2}M^{2}k^{2}-m^{2}M^{4})M_{V}^{2}\\
&+& (d-3)k^{2}m^{2}(k^{2}-M^{2})^{2}\biggr], \\
c_{MB}^{(\gamma)} &=& 
-\frac{2m}{(d-2)k^{2}(k^{2}-4m^{2})}\biggl[(M_{V}^{2}+(d-3)k^{2})((2m^{2}
-M^{2})k^{2}+2m^{2}(M^{2}-M_{V}^{2}))\biggr], \\
c_{MV}^{(\gamma)} &=& \frac{2m}{(d-2)(k^{2}-4m^{2})}
\biggl[(M_{V}^{2}+(d-3)k^{2})(k^{2}-M^{2}-M_{V}^{2})\biggr],
\end{eqnarray*}
and the coefficients $c^{(k)}$ are given by
\begin{eqnarray*}
c_{MBV}^{(k)} 
&=& \frac{2m^{2}}{(d-2)k^{4}(k^{2}-4m^{2})}\biggl[((d-2)k^{2}
-4(d-1)m^{2})M_{V}^{6} \\ 
&-& 2((d-2)k^{4}+(dM^{2}-2(d+1)m^{2})k^{2}-4(d-1)m^{2}M^{2})M_{V}^{4}\\ 
&+& ((d-2)k^{6}+4(d-5)m^{2}k^{4}-2(d-4)M^{2}k^{4}+((d-2)k^{2}
-4(d-1)m^{2})M^{4})M_{V}^{2}\\ &-& 4(d-3)k^{2}m^{2}(k^{2}-M^{2})^{2}\biggr], \\
c_{MB}^{(k)} &=&
\frac{1}{(d-2)k^{4}(k^{2}-4m^{2})}\biggl[(4(d-3)m^{2}(2m^{2}-M^{2})
-(d-2)(2m^{2}+M^{2})M_{V}^{2})k^{4} \\ 
&+&
2m^{2}((d-2)M_{V}^{4}+(d-4)M^{2}M_{V}^{2}+4m^{2}((d-3)M^{2}+2M_{V}^{2}))k^{2}
\\ 
&+& 8(d-1)m^{4}(M^{2}-M_{V}^{2})M_{V}^{2}\biggr],\\
c_{MV}^{(k)} 
&=& -\frac{4m^{2}}{(d-2)k^{2}(k^{2}-4m^{2})}\biggl[(M_{V}^{2}
+(d-3)k^{2})(k^{2}-M^{2}-M_{V}^{2})\biggr], \\
c_{M}^{(k)} &=& \frac{M_{V}^{2}}{k^{2}}.
\end{eqnarray*}
Here we used the abbreviations $k^{4}\equiv (k^{2})^{2}$ and 
$k^{6}\equiv (k^{2})^{3}$. In view of the denominators of the coefficients 
$c^{(\gamma,k)}$, which contain powers of $k^{2}$, it is advantageous to 
expand the scalar loop integrals in the small variable $\beta 
= \gamma\tilde\beta$ first. From eqs.~(\ref{eq:IMVIR},\ref{eq:IMBVIR}), 
we find
\begin{displaymath}
I_{MV}^{IR} = \frac{I_{M}}{m^{2}(\alpha-\gamma)}
+\beta\biggl(\frac{(d\gamma-(d-4)\alpha)I_{M}}{m^{2}d(\alpha-\gamma)^{3}}\biggr)+\ldots
\end{displaymath}
 and
\begin{eqnarray*}
I_{MBV}^{IR} 
&=& \frac{I_{MB}^{IR}}{m^{2}(\alpha-\gamma)}\\ 
&+& \beta\biggl(\frac{((d-1)(\gamma-\alpha)(\alpha-2)
-2(4\alpha-\alpha^{2}))m^{2}I_{MB}^{IR}+((d-3)\alpha
-(d-1)\gamma)I_{M}}{2(d-1)m^{4}(\gamma-\alpha)^{3}}\biggr)\\
&+& O(\beta^{2}).
\end{eqnarray*}
 Inserting the $\beta$-expansions of the scalar loop integrals in the
expressions for $I_{1+2}^{(\gamma,k)}$ from
eqs.~(\ref{eq:I12gdecomp},\ref{eq:I12kdecomp}), 
one observes that the poles in the variable $\beta\sim k^{2}$ cancel. 
In the final step, we must insert the expansion of the scalar loop integral 
$I_{MB}^{IR}$ in the second small variable $\alpha\sim M^{2}$, which can
directly be read off from eq.~(\ref{eq:IMBIR}). The expression for $I_{M}$ 
is given in eq.~(\ref{eq:IM}). Doing this, taking the limit $d\rightarrow 4$ 
and comparing to the decomposition 
of the matrix element in eq.~(\ref{eq:AxMatEl}), one finds the following 
$O(q^{3})$-contribution of $I_{1}$ and $I_{2}$ to the axial form factor $G_{A}$:
\begin{equation} \label{eq:resultGA}
G_{A}^{1+2} = -\frac{g_{1}}{3\pi}M^{3} + O(q^{4}) 
= -\frac{2\sqrt{2}g_{A}G_{V}R_{V}}{3\pi M_{V}^{2}F^{2}}M^{3} + O(q^{4}).
\end{equation}
There are no terms of lower order. This is in accord with the power 
counting for the two graphs, which predicts a chiral order of $q^{3}$ for 
this contribution to $G_{A}$. In order to compare this with the $q^{3}$-terms 
in $G_{A}$ worked out in \cite{Schweizer}, one can proceed as follows. 
Looking at fig.~\ref{fig:triangles}, and imagining the vector meson lines 
shrinking to a point vertex (corresponding to a limit where the mass $M_{V}$ 
tends to infinity, with $G_{V}R_{V}/M_{V}$ fixed), it is intuitively clear 
that the result corresponds to a pion-nucleon loop graph with an $O(q^{2})$ 
contact term replacing the vector meson line. Such contributions are 
parametrized by the two LECs $c_{3}$ and $c_{4}$ in \cite{Schweizer}, 
see eq.~(\ref{eq:GaSchweizer}). In fact, identifying
\begin{equation} \label{eq:c4}  
\frac{2\sqrt{2}G_{V}R_{V}}{M_{V}^{2}} = -c_{4},
\end{equation}
one reproduces exactly the corresponding terms in the representation based on
the pure pion-nucleon theory, cf.  eq.~(\ref{eq:GaSchweizer}).
That eq.~(\ref{eq:c4}) is a good guess can be seen like that: 
Comparing the $\rho N$-coupling used here, namely, the term proportional 
to $R_{V}$ in eq.~(\ref{eq:LWN}), to a more conventional one using a vector 
field representation for the rho field,
\begin{equation}
\mathcal{L}_{NV} = \frac{1}{2}g_{\rho NN}\bar\psi\biggl(\gamma^{\mu}
\mathbf{\rho_{\mu}\cdot\tau}-\frac{\kappa_{\rho}}{2m}
\sigma^{\mu\nu}\partial_{\nu}\mathbf{\rho_{\mu}\cdot\tau}\biggr)\psi,
\end{equation}
one deduces
\begin{displaymath}
g_{\rho NN}\kappa_{\rho} = -\frac{4\sqrt{2}mR_{V}}{M_{V}}.
\end{displaymath}
Using this in eq.~(\ref{eq:c4}), we get
\begin{displaymath}
c_{4} = \frac{g_{\rho NN}\kappa_{\rho}G_{V}}{2mM_{V}} = \frac{\kappa_{\rho}}{4m}.
\end{displaymath}
In the last step, we have assumed a universal rho coupling, 
$M_{V}G_{V} \equiv F^{2}g_{\rho\pi\pi} = F^{2}g_{\rho NN}$ as well as 
the KSFR relation $M_{V}^{2}=2F^{2}g_{\rho NN}^{2}$ \cite{KSFR66} (see also
the recent discussion in the framework of effective field theory in 
ref.~\cite{Djukanovic:2004mm}). 
This agrees with the rho-contribution to $c_{4}$ found in \cite{AspectsPiN}. 
Furthermore, there is no rho contribution to the LEC $c_{3}$ according to this
work.

The result of eq.~(\ref{eq:c4}) is not surprising for itself, but the
agreement of our findings with previous resonance saturation analyses 
demonstrates one very important thing, namely, that the variants of the 
infrared regularization scheme derived in the previous sections are consistent 
with the standard case of infrared regularization in the pion-nucleon sector 
used in \cite{Schweizer}. 

As a side remark, we note that the leading order result from the triangle 
graphs shows no $t-$dependence and therefore gives no contribution to the 
axial radius. However, the leading contribution to the axial radius can also 
be related to meson resonances, namely, to a tree-level exchange of an 
axial-vector meson. The pertinent calculation can be found in
\cite{Schindler}, where the axial vector meson-couplings to the pions and 
nucleons are fitted to experimental data for $G_{A}(t)$ (for an earlier
study based on chiral Lagrangians, see \cite{Gari:1984qs}).  Equivalently, 
it can be parametrized by a certain LEC, named $d_{22}$ in \cite{Schweizer} 
(see eq.~(\ref{eq:GaSchweizer})). As already mentioned at the beginning 
of sec.~\ref{sec:AxFF}, the difference between the two approaches just 
amounts to a resummation of higher order terms. Compared to the leading 
order term, the $t$-dependent part derived from the triangle graphs is 
suppressed by factors of the small variable $\alpha$, which is a reflection 
of the fact that the infrared regularized loop integrals preserve the chiral 
power counting.   

\section{Summary} \label{sec:summary}

In this paper we have presented an extension of the infrared regularization 
scheme that allows for an inclusion of explicit (vector and axial-vector) meson 
resonances in the single-nucleon sector of ChPT. For the processes we 
have considered here, the meson resonances do not appear as external
particles, and the corresponding power counting rules for the internal 
resonance lines are set up such that the resonance four-momentum is considered 
to be small compared to its mass. The infrared regularization scheme extracts 
the part of the one-loop graphs to which this power counting scheme applies 
(for any value of the dimension parameter $d$ used in dimensional
regularization), while the remaining parts of the loop graphs will in general 
violate the power counting requirements, but can be absorbed in a 
renormalization of the local terms of the effective Lagrangian. 

After a short review of the infrared regularization procedure used for 
the pion-nucleon and the vector meson-pion system in sec.~\ref{sec:PiN} 
and \ref{sec:PiV}, respectively, we have combined the analyses of these 
sections in sec.~\ref{sec:PiNV}. There, we consider the simplest example 
of a Feynman graph where nucleons, pions as well as (vector) meson resonances 
show up. It is shown how to extract the infrared singular part of such a
graph, and the power counting requirements are verified. It should be clear 
from this example how the infrared singular parts of more complicated one-loop 
graphs (with more nucleon, pion and resonance lines) can be worked out 
(some remarks on the general case can be found at the beginning of 
sec.~\ref{sec:PiNV}, and in sec.~6 of \cite{BL}). Finally, in 
sec.~\ref{sec:AxFF}, we have applied the extended scheme to compute a 
vector meson induced loop contribution to the axial form factor of the 
nucleon, and demonstrate that the result agrees with the result for $G_{A} (t)$ 
given by Schweizer \cite{Schweizer} in combination with the resonance 
saturation analysis for the pion-nucleon LECs in \cite{AspectsPiN}. 
There are, of course, many other possibilities for applications of the 
scheme developed here. Finally, we add the remark that a suggestion for 
an extension of infrared regularization to the multiloop case was made 
in ref.~\cite{LP}.

\acknowledgments{This research is part of the EU Integrated Infrastructure Initiative Hadron Physics
 Project under contract number RII3-CT-2004-506078. Work supported in part by
 DFG (SFB/TR 16, ``Subnuclear Structure of Matter'') and by the Helmholtz Association
 through funds provided to the virtual institute ``Spin and strong
 QCD'' (VH-VI-231).}

\begin{appendix}

\section{Explicit expression for {\boldmath$I_{MV}^{IR}$}}
\label{app:closedform}
\def\theequation{\Alph{section}.\arabic{equation}}
\setcounter{equation}{0}

In eq.~(8.4) of ref.~\cite{BM04}, a closed  expression for $I_{MV}^{IR}$ for
$d\rightarrow 4$ was given that looks much simpler than our result, 
eq.~(\ref{eq:IMVIR}), where there is still an infinite sum to be performed. 
On the other hand, it is quite tedious to work out the chiral expansion of 
the result of \cite{BM04}, due to the rather complicated expressions
\begin{equation} \label{eq:x1x2}
x_{1,2} = \frac{1}{2\tilde\beta}\biggl(\tilde\beta+\tilde\alpha 
-1 \pm \sqrt{(\tilde\beta+\tilde\alpha -1)^{2}-4\tilde\alpha\tilde\beta}\biggr)
\end{equation}
used there. These two expressions are nothing but the zeroes of the Feynman 
parameter integral encountered in the computation of the loop integral. 
The corresponding chiral expansions start with
\begin{eqnarray*}
x_{1} &=& -\tilde\alpha + \ldots , \\
x_{2}^{-1} &=& -\tilde\beta + \ldots ,
\end{eqnarray*}
showing that $x_{1}$ is small and negative, while $x_{2}$ tends to minus
infinity for $\tilde\beta\rightarrow 0_{+}$. In order to show the equivalence 
of the two results for $I_{MV}^{IR}$, we employ the following relations,
\begin{eqnarray}
\tilde\beta(x_{1}+x_{2}) &=& \tilde\beta+\tilde\alpha-1, \label{eq:sumx1x2} \\
\tilde\beta x_{1}x_{2} &=& \tilde\alpha , \label{eq:prodx1x2}
\end{eqnarray}
to write 
\begin{displaymath}
\frac{(\tilde\alpha\tilde\beta)^{k}}{(1-\tilde\alpha-\tilde\beta)^{2k+1}} 
= -\frac{1}{\tilde\beta
  x_{2}}\frac{(\frac{x_{1}}{x_{2}})^{k}}{(1+\frac{x_{1}}{x_{2}})^{2k+1}}= 
-\frac{1}{\tilde\beta x_{2}}\sum_{m=0}^{\infty}\frac{(-1)^{m}
\Gamma(2k+m+1)}{m!\Gamma(2k+1)}\biggl(\frac{x_{1}}{x_{2}}\biggr)^{k+m}.
\end{displaymath}
Inserting this in eq.~(\ref{eq:IMVIR}) yields
\begin{eqnarray} \label{eq:interstep}
I_{MV}^{IR}(q^{2}) 
&=& \frac{M_{V}^{d-4}(\tilde\alpha)^{\frac{d}{2}-1}}{(4\pi)^{\frac{d}{2}}
\tilde\beta x_{2}}\sum_{k=0}^{\infty}\sum_{m=0}^{\infty}\frac{\Gamma(2k+m+1)
\Gamma(1-\frac{d}{2}-k)\Gamma(2k+1)}{\Gamma(2k+1)\Gamma(k+1)\Gamma(m+1)}
\biggl(-\frac{x_{1}}{x_{2}}\biggr)^{k+m}  \nonumber \\
&=&
\frac{M_{V}^{d-4}(\tilde\alpha)^{\frac{d}{2}-2}x_{1}}{(4\pi)^{\frac{d}{2}}}
\sum_{k=0}^{\infty}\sum_{m=0}^{\infty}\frac{\Gamma(2k+m+1)\Gamma(1
-\frac{d}{2}-k)}{\Gamma(k+1)\Gamma(m+1)}\biggl(-\frac{x_{1}}{x_{2}}\biggr)^{k+m}. 
\end{eqnarray}
In the second line we made use of eq.~(\ref{eq:prodx1x2}). 
Now we change the summation indices according to
\begin{displaymath}
j = k+m, \qquad l = k = j-m,
\end{displaymath}
and use the following sum formula for Gamma functions,
\begin{equation} \label{eq:GammaSum}
\sum_{l=0}^{j} \frac{\Gamma(j+l+1)\Gamma(x-l)}{\Gamma(j-l+1)\Gamma(l+1)} 
= (-1)^{j}\frac{\Gamma(x-j)\Gamma(-x)}{\Gamma(-x-j)}, \qquad j\in\mathbf{N},
\end{equation}
for $x=1-\frac{d}{2}$. We shall give a short outline of a proof for
eq.~(\ref{eq:GammaSum}): Dividing this equation by $\Gamma(x-j)$, both sides
are just polynomials in $x$ of degree $j$, with coefficient $1$ in front of
$x^{j}$. Consequently, one only has to prove that both polynomials have the 
same set of zeroes, namely $\lbrace -1,-2,\ldots,-j\rbrace $. This is not
difficult, making use of
\begin{eqnarray*}
\sum_{l=i}^{j} \frac{(-1)^{j-l}\Gamma(j-i+1)}{\Gamma(j-l+1)
\Gamma(l+1)}\prod_{p=0}^{i-1}(l-p) &=& 
\sum_{l=i}^{j}\frac{(-1)^{j-l}\Gamma(j-i+1)}{\Gamma(j-l+1)\Gamma(l-i+1)} \\
= \sum_{n=0}^{j-i}{j-i \choose n}(-1)^{j-i-n} &=& (1-1)^{j-i} = 0 
\end{eqnarray*}
for $0 < i < j$. These remarks should be sufficient to complete the proof 
of eq.~(\ref{eq:GammaSum}).

Returning to eq.~(\ref{eq:interstep}), we employ eq.~(\ref{eq:GammaSum}) to write
\begin{eqnarray} \label{eq:resultIMVsoft}
I_{MV}^{IR}(q^{2}) 
&=&
\frac{M_{V}^{d-4}(\tilde\alpha)^{\frac{d}{2}-2}x_{1}}{(4\pi)^{\frac{d}{2}}}
\sum_{j=0}^{\infty}\sum_{l=0}^{j}\frac{\Gamma(j+l+1)\Gamma(1
-\frac{d}{2}-l)}{\Gamma(j-l+1)\Gamma(l+1)}\biggl(-\frac{x_{1}}{x_{2}}\biggr)^{j} 
\nonumber \\
&=&
\frac{M_{V}^{d-4}(\tilde\alpha)^{\frac{d}{2}-2}x_{1}}{(4\pi)^{\frac{d}{2}}}
\sum_{j=0}^{\infty}\frac{\Gamma(\frac{d}{2}-1)\Gamma(1
-\frac{d}{2}-j)}{\Gamma(\frac{d}{2}-1-j)}\biggl(\frac{x_{1}}{x_{2}}\biggr)^{j}.  
\end{eqnarray}
The last line of eq.~(\ref{eq:resultIMVsoft}) is exactly the result that was
derived in sections~7 and 8 of \cite{BM04}. This can be seen by substituting 
\begin{displaymath}
a\rightarrow -x_{1},\quad b\rightarrow -x_{2}^{-1},\quad  
d\rightarrow \frac{d}{2}-2
\end{displaymath}
in eq.~(7.7) of that reference, and multiplying the result with
\begin{displaymath}
-\frac{\Gamma(2-\frac{d}{2})M_{V}^{d-4}}{(4\pi)^{\frac{d}{2}}}(-\tilde\beta x_{2})^{\frac{d}{2}-2}\quad,
\end{displaymath}
as explained at the beginning of sec.~8 of \cite{BM04}. In the limit
$d\rightarrow 4$, 
the series of eq.~(\ref{eq:resultIMVsoft}) can be summed up to give
\begin{equation} \label{eq:IMVd4}
I_{MV}^{IR}(q^{2},d\rightarrow 4) = 2x_{1}\lambda-\frac{1}{16\pi^{2}}
\biggl(x_{1}(1-\ln\tilde\alpha)-(x_{1}-x_{2})\ln\biggl(1-\frac{x_{1}}{x_{2}}
\biggr)\biggr),
\end{equation}
where
\begin{displaymath}
\lambda = \frac{M_{V}^{d-4}}{16\pi^{2}}\biggl(\frac{1}{d-4}-
\frac{1}{2}(\ln(4\pi)-\gamma+1)\biggr).
\end{displaymath}
Eq.~(\ref{eq:IMVd4}) is identical to eq.~(8.4) of \cite{BM04}.

\section{Alternative derivation of {\boldmath$I_{MBV}^{IR}$} }
\label{app:TangIR}
\def\theequation{\Alph{section}.\arabic{equation}}
\setcounter{equation}{0}

In this appendix, we present an alternative derivation of the infrared
singular part of the loop integral $I_{MBV}(k^{2})$ (see
eqs.~(\ref{eq:IMBV},\ref{eq:IMBVIR})) using the prescription of Ellis and Tang 
we have already explained at the end of sec.~\ref{sec:PiV}. However, at some
places we will also use Feynman parameter integrals, so the derivation
outlined here is to some extent a mixture of the standard infrared
regularization procedure and the method of Ellis and Tang. In complete analogy 
to the steps performed at the end of sec.~\ref{sec:PiV}, we start with  
\begin{eqnarray}\label{eq:Tangsteps}
I_{MBV}(k^{2}) &\rightarrow& 
\int\frac{d^{d}l}{(2\pi)^{d}}\frac{i}{((p-l)^{2}-m^{2})(l^{2}-M^{2})}
\sum_{j=0}^{\infty}\frac{(-2k\cdot l)^{j}}{(k^{2}+l^{2}-M_{V}^{2})^{j+1}} \nonumber\\
&\rightarrow&
\int\frac{d^{d}l}{(2\pi)^{d}}\frac{i}{((p-l)^{2}-m^{2})(l^{2}-M^{2})}
\sum_{j=0}^{\infty}\frac{(-2k\cdot l)^{j}}{(k^{2}+M^{2}-M_{V}^{2})^{j+1}} \nonumber \\ 
&\rightarrow& \sum_{j=0}^{\infty}\int\frac{d^{d}l}{(2\pi)^{d}}
\frac{i(-2k\cdot l)^{j}}{(k^{2}+M^{2}-M_{V}^{2})^{j+1}((p-l)^{2}-m^{2})(l^{2}-M^{2})} .
\end{eqnarray}
Now we could use the procedure outlined in \cite{Tang} to expand the nucleon 
propagator, together with an interchange of summation and integration. In the 
present case, however, it is easy to see that it is equivalent to use the
common Feynman parameter trick for the remaining loop integrals {\em and} 
extend the parameter integration to infinity like in sec.~\ref{sec:PiN}. 
For $I_{MB}$ (see eqs.~(\ref{eq:IMB}) and (\ref{eq:intlz})), the splitting 
of eq.~(\ref{eq:split}) corresponds to 
\begin{equation} \label{eq:Tangsplit}
\frac{1}{((p-l)^{2}-m^{2})(l^{2}-M^{2})} 
= \frac{1}{p^{2}-m^{2}-2l\cdot p+M^{2}}\biggl(\frac{1}{l^{2}-M^{2}}
-\frac{1}{(p-l)^{2}-m^{2}}\biggr).
\end{equation}
This has also been noted in \cite{BL}, see eqs.~(22,23) of that reference. 
On the other hand, the first term on the r.h.s of eq.~(\ref{eq:Tangsplit}) is 
exactly the integrand that gives the soft momentum contribution in the sense
of Ellis and Tang, see e.g. eq.~(7) in \cite{Tang}. Therefore, it is
consistent to continue the series of steps in eq.~(\ref{eq:Tangsteps}) with
\begin{eqnarray} \label{eq:nextTangstep}
\ldots &\rightarrow& 
\sum_{j=0}^{\infty}\int_{0}^{\infty}\frac{dz}{(k^{2}+M^{2}-M_{V}^{2})^{j+1}}
\int\frac{d^{d}l}{(2\pi)^{d}}\frac{i(-2k\cdot l)^{j}}{[((p-l)^{2}
-m^{2})z+(l^{2}-M^{2})(1-z)]^{2}} \nonumber \\ 
&\equiv& I_{MBV}^{\mathrm{soft}}(k^{2}).
\end{eqnarray}
The loop-integration can be done using the following generalization 
of eq.~(\ref{eq:supertadpole}):
\begin{equation} \label{eq:megatadpole}
\int\frac{d^{d}l}{(2\pi)^{d}}\frac{i(k\cdot l)^{2n}}{(l^{2}-M^{2})^{r}} 
= (-1)^{r-1}(-k^{2}M^{2})^{n}M^{d-2r}\frac{\Gamma(n+\frac{1}{2})}{
\Gamma(\frac{1}{2})}\frac{\Gamma(r-\frac{d}{2}-n)}{(4\pi)^{\frac{d}{2}}\Gamma(r)}  
\end{equation} 
for $r,n \in \mathbf{N}$. This gives 
\begin{displaymath}
I_{MVB}^{\mathrm{soft}}(k^{2}) 
= \sum_{j=0}^{\infty}\sum_{i=0,i\in 2\mathbf{N}}^{j}
\frac{m^{d-6}2^{i}{j\choose
    i}(-\beta)^{j-\frac{i}{2}}}{(\gamma-\alpha-\beta)^{j+1}}
\frac{\Gamma(\frac{i+1}{2})\Gamma(2-\frac{d}{2}-\frac{i}{2})}{
\Gamma(\frac{1}{2})(4\pi)^{\frac{d}{2}}}\int_{0}^{\infty}\frac{dzz^{j-i}}{
(z^{2}-\alpha z+\alpha)^{2-\frac{d}{2}-\frac{i}{2}}}.
\end{displaymath} 
Here we have also used the on-shell kinematics specified in
eq.~(\ref{eq:kinppbar}). 
The sum over the even integers $i$ extends only to $j-1$ if $j$ is odd. 
Defining new indices, 
\begin{displaymath}
J = j-\frac{i}{2}, \qquad l=\frac{i}{2},
\end{displaymath}
and reordering the series correspondingly gives the following expression 
for $I_{MBV}^{\mathrm{soft}}(k^{2})$:
\begin{displaymath}
 \sum_{J=0}^{\infty}\sum_{l=0}^{J}\frac{m^{d-6}(-\beta)^{J}4^{l}
\Gamma(J+l+1)\Gamma(l+\frac{1}{2})\Gamma(2-\frac{d}{2}-l)}{(4\pi)^{\frac{d}{2}}
(\gamma-\alpha-\beta)^{J+l+1}\Gamma(J-l+1)
\Gamma(2l+1)\Gamma(\frac{1}{2})}\int_{0}^{\infty}
\frac{z^{J-l}dz}{(z^{2}-\alpha z+\alpha)^{2-\frac{d}{2}-l}}.
\end{displaymath}
Using eq.~(\ref{eq:GammaId2}), it is straightforward to see that this 
equals $I_{MBV}^{IR}$ of eq.~(\ref{eq:IMBVIR}), as expected (the remaining 
parameter integral can be done with the help of eq.~(\ref{eq:yparint})).

\section{Loop Integrals}
\label{app:LoopInt}
\def\theequation{\Alph{section}.\arabic{equation}}
\setcounter{equation}{0}
Here we list the decomposition of the loop integrals with tensor structures in 
the numerator, which we need in sec.~\ref{sec:AxFF}. All loop integrals in 
this appendix are understood as the infrared singular parts of the full loop 
integrals, but we will suppress the superscript $IR$ for brevity. As a
consequence, all loop integrals that do not contain a pion propagator are
already dropped here, since they have no infrared singular part. Also, we will 
use the mass shell condition $p^{2}=m^{2}$ for the nucleon momentum $p$. We start with
\begin{equation}
\int\frac{d^{d}l}{(2\pi)^{d}}\frac{il^{\mu}}{((p-l)^{2}-m^{2})(l^{2}-M^{2})} 
= p^{\mu}I^{(1)}_{MB},
\end{equation}
where
\begin{displaymath}
I^{(1)}_{MB} = \frac{1}{2m^{2}}\biggl(M^{2}I_{MB}-I_{M}\biggr).
\end{displaymath}
In complete analogy,
\begin{equation}
\int\frac{d^{d}l}{(2\pi)^{d}}\frac{il^{\mu}}{((l-k)^{2}-M_{V}^{2})(l^{2}-M^{2})} 
= k^{\mu}I^{(1)}_{MV},
\end{equation}
with
\begin{displaymath}
I^{(1)}_{MV} = \frac{1}{2k^{2}}\biggl((k^{2}+M^{2}-M_{V}^{2})I_{MV}-I_{M}\biggr).
\end{displaymath}
Integrals of type $MB$ and $MV$ are also needed with a tensor structure 
$l^{\mu}l^{\nu}$ in the numerator. They are decomposed as
\begin{equation}
\int\frac{d^{d}l}{(2\pi)^{d}}\frac{il^{\mu}l^{\nu}}{((l-k)^{2}-M_{V}^{2})(l^{2}-M^{2})} 
= g^{\mu\nu}t^{(0)}_{MV}(k)+\frac{k^{\mu}k^{\nu}}{k^{2}}t^{(1)}_{MV}(k),
\end{equation}
where the coefficients of the tensor structures are given by
\begin{eqnarray*}
(d-1)t^{(0)}_{MV}(k) &=& 
\frac{4k^{2}M^{2}-(k^{2}+M^{2}-M_{V}^{2})^{2}}{4k^{2}}I_{MV}
+\frac{k^{2}+M^{2}-M_{V}^{2}}{4k^{2}}I_{M}, \\
(d-1)t^{(1)}_{MV}(k) &=& 
\frac{d(k^{2}+M^{2}-M_{V}^{2})^{2}-4k^{2}M^{2}}{4k^{2}}I_{MV}-
\frac{d(k^{2}+M^{2}-M_{V}^{2})}{4k^{2}}I_{M}.
\end{eqnarray*}
The corresponding coefficients in the meson-baryon case, $t_{MB}^{(0,1)}(p)$, 
can be derived from these results by substituting $k\rightarrow p,
M_{V}\rightarrow m$.

We turn now to loop integrals with three propagators. First the vector integral:
\begin{equation}
\int\frac{d^{d}l}{(2\pi)^{d}}\frac{il^{\mu}}{((\bar p-l)^{2}
-m^{2})((l-k)^{2}-M_{V}^{2})(l^{2}-M^{2})} 
= (k+\bar p)^{\mu}I_{MBV}^{A}+(k-\bar p)^{\mu}I_{MBV}^{B},
\end{equation}
with
\begin{eqnarray*}
I_{MBV}^{A} &=& \frac{1}{2k^{2}(4m^{2}-k^{2})}\biggl[(2m^{2}M^{2}
+(2m^{2}-k^{2})(k^{2}-M_{V}^{2}))I_{MBV} \\ &-& k^{2}I_{MV}-(2m^{2}-k^{2})I_{MB}\biggr], \\
I_{MBV}^{B} &=&
\frac{1}{2k^{2}(4m^{2}-k^{2})}\biggl[(2M^{2}(m^{2}-k^{2})+(k^{2}
-M_{V}^{2})(k^{2}+2m^{2}))I_{MBV} \\ &+&3k^{2}I_{MV}-(k^{2}+2m^{2})I_{MB}\biggr].
\end{eqnarray*}
We remind the reader that we use $\bar p^{2}=m^{2}=(\bar p-k)^{2}$ here. The
scalar loop integral with three propagators occuring in this decomposition 
was named $\tilde I_{MBV}$ in sec.~\ref{sec:PiNV}, eq.~(\ref{eq:IMBV2}). 
However, we noted there that it is equal to $I_{MBV}$ for on-shell nucleon
momenta. 

The tensor integral
\begin{equation}
I_{MBV}^{\mu\nu} 
= \int\frac{d^{d}l}{(2\pi)^{d}}\frac{il^{\mu}l^{\nu}}{((\bar p-l)^{2}
-m^{2})((l-k)^{2}-M_{V}^{2})(l^{2}-M^{2})}
\end{equation}
can be decomposed as
\begin{displaymath}
I_{MBV}^{\mu\nu}= g^{\mu\nu}C_{1} + (\bar p+k)^{\mu}(\bar p+k)^{\nu}C_{2}
+(\bar p-k)^{\mu}(\bar p-k)^{\nu}C_{3}
+((\bar p+k)^{\mu}(k-\bar p)^{\nu}+(\bar p+k)^{\nu}(k-\bar p)^{\mu})C_{4}.
\end{displaymath}
Here the coefficients $C_{i}$ are given by 
\begin{eqnarray*}
C_{1} &=& 
\frac{1}{d-2}\biggl[M^{2}I_{MBV}-\frac{1}{2}(k^{2}-M_{V}^{2}+2M^{2})I_{MBV}^{A}
+\frac{M_{V}^{2}-k^{2}}{2}I_{MBV}^{B}\biggr], \\
C_{2} &=& \frac{1}{k^{2}(4m^{2}-k^{2})}
\biggl[m^{2}M^{2}I_{MBV}+\frac{M_{V}^{2}-k^{2}}{2}\biggl(m^{2}I_{MBV}^{B}
+(k^{2}-m^{2})I_{MBV}^{A}\biggr) \\ 
&-& \frac{2m^{2}-k^{2}}{4}I_{MB}^{(1)} - \frac{k^{2}}{4}I_{MV}^{(1)} 
- m^{2}(d-1)C_{1}\biggr],\\
C_{3} &=& \frac{1}{k^{2}(4m^{2}-k^{2})}\biggl[M^{2}(m^{2}+2k^{2})I_{MBV}
-\frac{1}{2}(k^{2}-M_{V}^{2}+2M^{2})((m^{2}+2k^{2})I_{MBV}^{A} \\ 
&+& (k^{2}-m^{2})I_{MBV}^{B})+\frac{3k^{2}}{4}I_{MV}^{(1)}+
\frac{2m^{2}+k^{2}}{4}I_{MB}^{(1)}-(m^{2}+2k^{2})(d-1)C_{1}\biggr], \\
C_{4} &=& 
\frac{1}{k^{2}(4m^{2}-k^{2})}\biggl[M^{2}(m^{2}-k^{2})I_{MBV}
+\frac{k^{2}-M_{V}^{2}}{2}\biggl((2k^{2}+m^{2})I_{MBV}^{A}\\
&+& (k^{2}-m^{2})I_{MBV}^{B}\biggr)  
+ \frac{3k^{2}}{4}I_{MV}^{(1)}-\frac{2m^{2}+k^{2}}{4}I_{MB}^{(1)}\\
&-& (m^{2}-k^{2})(d-1)C_{1}\biggr].
\end{eqnarray*}

\end{appendix}

\end{document}